# Singular-Turbulent Structure Formation in the Universe and the Essence of Dark Matter I. Unified model for dark matter and quintessence

Reza Dastvan

*Institute of Biochemistry and Biophysics, University of Tehran, P.O. Box 13145-1384, Tehran, Iran.* <u>dastvan@khayam.ut.ac.ir</u>

ABSTRACT

Based on superfluid behavior of a (boson) dark matter as the light itself, a unified model for dark matter and quintessence is proposed. Inspired by (O'Dell *et al.* 2000) which in an exciting study showed that particular configurations of intense off-resonant laser beams (simulate the Bose-Einstein condensate) can give rise to an attractive $1/r$ interatomic potential, I infer electromagnetic origin of inertia and gravity. Time varying dark components result in time varying electromagnetic fine structure constant ($\alpha$) or the quantities given in its definition (even an invariant $\alpha$ would not exclude the possibility that these values might be changing in ratio to one another), in either case variable speed of light (VSL) with $\dot{c}<0$. According to (Wetterich 2003a) a bound $|R| = |\Delta\alpha(z = 0.13)/\Delta\alpha(z = 2)| < 0.02$ strongly favors quintessence with a time varying equation of state $w = p/\rho$, where the value of $(1+w)$ at present is substantially smaller than for $z = 2$ and equivalently compressibility is increased in the medium and this process has significant influences on the process of structure formation. It is possible that this monotonically increasing trend of $w$ extends to high-redshifts. These models typically require quintessence coupling to a photon which would require large coupling between non-baryonic dark matter and the scalar field therefore consistent with our model, quintessence may couple to both non-baryonic dark matter and radiation. Wetterich (2003b) observe that $\Omega_{quint}$ does not affect the time evolution of the fundamental couplings and $\alpha$ depends only on the equation of state $w$ and this fascinating result further support the common essence of light and boson dark matter.

**Key words**: Cosmology: Structure Formation, Dark Matter, Variable Dark Energy, Superfluidity

*"All things are connected."* Chief Seattle (1854)

# 1 Introduction

The possible discrepancies between theory and observation have motivated new proposals for the nature of dark matter. Each proposed variation from standard cold, collisionless dark matter (SCDM) has two properties: (1) it can "solve" some or all of the problems described in the subsequent section, and (2) it leads to additional predictions that would distinguish it from all the other alternatives.

The hierarchical gravitational collapse of cold collisionless particles leads to dense, singular dark matter halos – a result that is central to several fundamental problems with this model on small scales. Self-interacting dark matter has been proposed to alleviate some of the problems of galaxy formation arising in the standard cold dark matter models (Spergel & Steinhardt 2000), (Wandelt *et al.* 2001). The Spergel-Steinhardt proposal has stimulated the interesting possibility that dark matter consists of particles that interact through the strong force with ordinary matter.

Since all theories unifying gravity with other interactions involve scalar fields, more "interesting" scenarios consider dark matter in the form of a scalar field coherent on a very large scale, similar to those associated with quintessence sources. Even if, until now, these remnants of primordial scalar fields have not been directly detected, their use as models for quintessence or dark matter in galactic halo structures has become widespread (Hernández *et al.* 2004). I propose a unified theory for quintessence and dark matter in galactic halos based on bosonic scalar field dark matter (Nishiyama, Morita & Morikawa 2004; Fukuyama & Morikawa 2006; Mangano, Miele & Pettorino 2003; Sahni & Wang 2000) and vacuum superfluid (Winterberg 2002; Volovik 2002; Guzman 2001; González-Díaz 2003). So far, most of the attempts to model galactic dark matter halos out of real or complex scalar fields assume that each galactic halo is a spherical Bose-Einstein condensate of scalar particles. Subsequent studies added self-interaction and generalized the previous Newtonian analysis to be fully general relativistic. Other authors contributed more detailed studies which used two kinds of coupling of the scalar field to gravity, i.e. either minimal or non-minimal.

Various bosonic and scalar field dark matter and/or quintessence candidates with different mass and time of transition to superfluid (BEC) state (at $z > z_d$ or LTPT) are proposed in literature (Ratra and Peebles 1988; Press et al. 1990; Schramm 1990; Gradwohl 1991a, b; Frieman, Hill & Watkins 1992; Sin 1994; Ji & Sin 1994; Lee & Koh 1996; Larsen and Madsen 1996; Hu, Barkana & Gruzinov 2000; Goodman 2000; Sahni & Wang 2000; Matos & Ureña-López 2000; Peebles 2000; Dymnikova & Khlopov 2000; Arbey, Lesgourgues & Salati 2001; Guzman 2001; Silverman & Mallett 2001a, b, 2002; Volovik 2002; Winterberg 2002; Yu & Morgan 2002; Alcubierre et al. 2002; Mangano, Miele & Pettorino 2003; Guzmán & Ureña-López 2003; Arbey, Lesgourgues & Salati 2003; Christianto 2003; Nishiyama, Morita & Morikawa 2004; Hernández et al. 2004; Ferrer and Grifols 2004; Amendola & Barbieri 2005; Guzmán & Ureña-López 2006; Fukuyama & Morikawa 2006). Although the alternative models could very well lead to particle masses different by several orders of magnitude, but that would not affect the principal conclusion - the possible existence of a Bose–Einstein condensate of cosmic extent - since this phenomenon could occur throughout a very wide range of masses. For example, particles with masses as large as $10^{-6}$ eV − which are still lower than any hypothesized for massive neutrinos − would form a Bose–Einstein condensate at critical temperatures below 1000 K (Silverman & Mallett 2001a).

Most of literature based on theoretical calculations and galactic information (rotation curves, luminous core size of galaxies, mean mass density, etc.) proposed an ultra low or low mass boson (Silverman & Mallett 2001b, 2002; Hu, Barkana & Gruzinov 2000; Sin 1994; Frieman, Hill & Watkins 1992; Sahni & Wang 2000; Matos & Ureña-López 2000; Larsen & Madsen 1996) and the boson mass estimated in most of these approaches roughly coincides with the value ~ 1.1-2× $10^{-23}$ −$10^{-22}$ eV/$c^2$. (Nishiyama, Morita & Morikawa 2004; Fukuyama & Morikawa 2006) proposed the upper limit of the boson mass < 2 eV and proposed that if the boson temperature is smaller than the radiation temperature, we cannot set the upper limit of the boson mass. One of the proposed boson candidates is ultra light pseudo Nambu-Goldstone boson (Schramm 1990; Sin 1994; Ji & sin 1994; Frieman et al. 1995; Yu & Morgan 2002). Although the others proposed massive scalar fields (Goodman 2000; Peebles 2000; Arbey et al. 2003) or Planck mass Bose particles (Winterberg 1992, 2002).

In these two papers, I aim to reconstruct the jigsaw puzzle of structure formation in the universe and I propose the outline of a comprehensive and unifying theory of dark matter and quintessence leaning on superfluid or Bose-Einstein condensate essence of dark matter. In this immense jigsaw puzzle, some fragments are not discovered yet but the frame of the scenario is proposed. In the subsequent papers several aspects of the scenario are discussed. In section 2, I describe a unified theory for quintessence and dark matter in galactic halos and the nature of dark matter. In section 3, based on the latest progresses in theoretical physics, new insights into the essences of wave-particle duality of matter, light, gravity and inertia, SCQGP, and CMB have proposed. In the final section, we investigate the relevancy of time varying constants with time varying dark components.

## 2  Unified model for dark matter & quintessence and superfluid behavior of DM

2.1 bosonic scalar field dark matter

Basic model we construct would have the following general features:

1. Boson dark matter or self-interacting (the dark matter interacts strongly with ordinary baryonic matter, as well as with itself) normal part (excitations) of superfluid boson dark matter, which possesses some nonvanishing amount of viscosity and friction is introduced as dark matter which initially dominates the energy density and in literature, the condensate (superfluid) part of the superfluid boson dark matter is identified as quintessence and SFDM (scalar field dark matter, equivalent of ΛCDM; Alcubierre *et al.* 2002) or FCDM (fuzzy cold dark matter (Hu, Barkana & Gruzinov 2000), a similar stuff of axion dark matter[*]).
2. One of the major differences of my model to the others is the nature of the observed universe acceleration. In this model, there is no obligation about the negative pressure of the quintessence which is a liquid in BEC state. In section 4, I introduce my convincing reasons about this exciting and adventuresome feature of the dark component.
3. When the energy density of BEC exceeds some critical value around $z = z_c$, the inhomogeneous components of the condensate would

---

[*] An ultra light pseudo-Goldstone boson which is cold by virtue of being born in a Bose condensate.

rapidly collapse on the turbulent coherent structures (maybe into massive object made of scalar field particles in quantum coherent states like boson stars or black holes) which recognized as SFDM in literature (Alcubierre *et al.* 2002, Silverman & Mallett 2002; Yu & Morgan 2002) and work as the standard cold dark matter halo of galaxies with the equation of state $p = 0$ (Fukuyama & Morikawa 2006; Nishiyama, Morita & Morikawa 2004; Yu & Morgan 2002; Alcubierre *et al.* 2002). New observations show that supermassive central objects lying in active galactic nuclei seem to be related to the velocity dispersion of the dark matter composing the dark halo, suggesting therefore that the central object was formed at the same time than the halo (Ferrarese & Merritt 2000; Gebhardt *et al.* 2000). This is probably contrary to the standard idea about galactic nuclei that proposes the existence of a central black hole, since it should be formed a time after the disc, *i.e.* much more after the formation of the halo, therefore the last word has not been said. These rapid collapses take place well after the decoupling stage, and therefore observed pattern in CMB fluctuations would not strongly be violated. I have developed a unified cosmological model of quintessence and DM using the Bose-Einstein condensation of a light boson and further the mechanism for the early formation of nonlinear objects. If we adopt this scenario, the quintessence dominance means that our universe is in almost the ground state described by a macroscopic wave function. Thus the largest scale of the Universe, as well as the smallest, would turn out to be described by quantum mechanics.

I introduce the following simplest model for the BEC: SFDM is identified as CDM with the equation of state $p = 0$ and the quintessence with $p = <w> \rho$ ($w \geq 0$). Although in this manner, the quintessence doesn't have negative pressure and it's not dark energy at all. I prefer to use its primary name and not to add a new member in essence family.

Let us consider the non-linear stage of the condensate collapse. I suppose a uniform spherically distributed region of BEC of the radius *r*. Self gravity of the sphere would further accelerate the collapse especially in the later stage. The collapse would continue until the Heisenberg uncertainty principle begins to support the structure, or a black hole is formed. Not all fluctuations will collapse into stable objects. Anyway the collapsed condensate forms localized compact objects classified as cold dark matter. Cold dark matter made up of a condensate of light particles would be frustrated in its attempts to cluster on small scales because of the uncertainty principle, the resulting SFDM (equivalent of FCDM, Sahni & Wang 2000; Hu, Barkana & Gruzinov 2000) might provide a natural explanation for two major difficulties faced by the standard CDM scenario. (i) The dearth of halo dwarf galaxies: the number of dwarf's in the local group is an order of magnitude smaller than predicted by N-body simulations of SCDM. (ii) The discrepancy between observed shapes of galaxy rotation curves and simulated dark matter halos. Recent observations of low surface brightness (LSB) galaxies show them to possess rotation curves which indicate a constant mass density in the central core region. These observations are difficult to accommodate within the SCDM model since high resolution N-body simulations of SCDM halos indicate a cuspy central density profile having the form $\rho \propto r^{-1.5}$ in the core region. SFDM inhibits gravitational clustering on small scales and could provide a natural resolution to the core density problem for disc galaxy halos. As a consequence of the quantum uncertainty principle, the particles of a BEC cannot be localized to regions smaller than the condensate coherence length (as determined by equilibrium between quantum pressure and gravitational attraction and this length is equivalent to the Jean's scale $\lambda_J$ for the onset of gravitationally unstable mass density perturbations) which, for particles of sufficiently low mass, corresponds to a size of the scale of the luminous core of galaxies. In this way SFDM within a galactic halo can provide the non-luminous mass needed to keep the galaxy together, yet not give rise to spike-like structures in the core or an excessive number of satellite structures (Alcubierre et al. 2002; Silverman & Mallett 2001a, b, 2002; Sahni & Wang 2000; Matos & Ureña-López 2000).

Based on Silverman & Mallett (2002), some galactic properties are: Condensate coherence length $\xi_C$,

$$\xi_c = \frac{h^2}{GMm^2} = \left(\frac{3h^2}{4\pi Gm^2 \bar{\rho}}\right)^{\frac{1}{4}} = \left(\frac{6\lambda_c^2}{\Lambda_c}\right)^{\frac{1}{4}} \tag{1}$$

determined by the boson mass, $m$, and condensate mass, $M$, irrespective of the radial variation in density. The second equality in equation (1) expresses $\xi_C$ in terms of the mean density defined by $M = \frac{4\pi}{3}\xi_C^3 \bar{\rho}$; the third equality expresses $\xi_C$ in terms of the boson Compton wavelength $\lambda_c$ and the condensate density parameter $\Lambda_c \equiv \frac{8\pi G \bar{\rho}}{c^2}$ (which differs from the cosmological constant $\Lambda$ since $\tilde{\rho} >> \rho_c$).

Quantum Jeans scale,

$$\lambda_{JQ} \sim \left( \frac{h^2}{G\bar{\rho}m^2} \right)^{1/4} \quad (2)$$

this corresponds to within a numerical factor of order unity to the coherence length $\xi_C$. Thus, density perturbations of a size less than $\xi_C$ are gravitationally *stable*. For fluctuations of a size much greater than $\xi_C$, SFDM behaves like ΛCDM.

Theoretical slightly increasing rotation curve of luminous matter,

$$v(u) = v_\infty \sqrt{1 - \frac{\tanh u}{u}} \quad (3)$$

SFDM is more like a wave than a particle, and the galactic halos are giant systems of condensed bose liquid. Three predictions are made:

(1) Mass profile $\rho \sim r^{-1.6}$ (Sin 1994);

(2) There are ripple-like fine structures in rotation curve (Sin 1994; Yu & Morgan 2002);

(3) As is well known, the bulk of a stationary superfluid, in contrast to a normal fluid, will remain stationary when its container is rotated. However, if a sample of superfluid of size $R$ is rotated at angular frequency $\omega$ which exceeds a critical frequency, then localized vortices can form. Yu & Morgan (2002) studied vortices in a rotating dark matter condensate comprehensively and showed that the presence of vortices is consistent with observed rotation curves of galaxies although even in the presence of vortices, the rotational motion of superfluid dark matter vortices would not show up as red- and blue-shifted subgalactic regions. Evidence of dark matter vortices, however, could conceivably be sought in rotationally-induced frame-dragging effects manifested through gravitational lensing or variation in polarization of transmitted light from distant background sources.

We now turn our attention to the global evolution of quintessence/DM in the expanding Universe. The evolution of the various energy densities is governed by the set of equations (the energy density of the quintessence part falls off and an expansion term must be subtracted from $\dot{\rho}_{quint}$),

$$\rho = \rho_{quint} + \rho_n + \rho_l, \ H \equiv \frac{\dot{a}}{a} = \sqrt{\frac{8\pi G\rho}{3}}, \ \rho_{BEC} = \rho_{quint} + \rho_l \quad (4)$$

$$\dot{\rho}_{quint} = \Gamma\rho_n - 3H\rho_{quint} - \Gamma'\rho_{quint}, \ \dot{\rho}_n = -3H\rho_n - \Gamma\rho_n, \ \dot{\rho}_l = -3H\rho_l + \Gamma'\rho_{quint},$$

$$\Gamma = \Gamma(z)$$

where $\rho_{quint}$, $\rho_l$ and $\rho_n$ are the energy densities of quintessence part of the BEC, localized energy density of SFDM (~ΛCDM) part of the BEC after the collapse and excited part of DM (which dominates initially), respectively. The sedimentation of BEC in the expanding Universe slowly proceeds with the time scale $\Gamma^{-1}$ but the rate of superfluid and BEC density increment is decreased along with aging of the universe, therefore the $\Gamma$ is a dynamic value. Nishiyama, Morita & Morikawa (2004) proposed that cycle of gradual sedimentation (transition of excited part to BEC) and rapid collapse of condensate (to SFDM) repeats many times and self-regularizes the ratio of quintessence and dark matter to be order one. But in this model, after the formation of cosmic coherent structures (described in the second paper) because of the matter flow induced by fluid dynamics of the cosmic foam, the condensation of boson dark matter to BEC proceeds continuously with the time scale $\Gamma^{-1}$ and above a critical $\rho_{BEC}$, a part of BEC rapidly collapses on coherent structures to form SFDM halo and central object of galaxies as described above. In addition to the *z*-dependent reduction in the rate of sedimentation which determines the trend of quintessence density increment, based on the so-called *Landau critical velocity criterion*, anywhere the relative velocity $|v_{quint} - v_n|$ of the coherent flow of the excitations (excited part of DM) in the cosmic foam with respect to the vacuum (quintessence part of the BEC) is larger than the Landau critical velocity, $|v_{quint} - v_n| > v_{Landau}$, new excitations (excited part of DM) can be created from the superfluid vacuum, therefore the vacuum is reconstructed due to created quasiparticles but remains superfluid (non-dissipative) with reduced density. These two processes regularize the ratio of quintessence and dark matter.

For the scalar $\varphi$-field to remain subdominant until recently its energy density must decrease rapidly at early times. Such behavior clearly cannot arise for polynomial potentials $V(\varphi) \propto \varphi^p$, $0 < p < few$, for which $\rho_{quint}$ will rapidly dominate the total density resulting in a colossal $\Lambda$-term today if $\rho_{quint} \sim \rho_{rad}$ initially.

Bassett *et al.* (2000) proposed a model of dark-matter candidates in which, if $V(\phi)$ has a quadratic minimum, the $\phi$-field will oscillate about this minimum and its average equation of state will be that of dust. This implies that the $\phi$-field will behave like axions or cold dark matter. Similarly if the potential is quartic, the average equation of state will be that of radiation.

An interesting potential which interpolates between an exponential and a power law is:

$$V(\Phi) = V_0 [\cosh(\lambda\Phi) - 1]^p \quad \text{(Sahni \& Wang 2000; Matos \& Ureña-López 2000)} \tag{5}$$

where $\lambda$ and *p* are constants, has the property that the scalar field equation of state mimics background matter $w \simeq w_B$ at early times whereas $<w> =$ ($p − 1)/(p + 1)$ at late times. Consequently, SFDM and maybe quintessence (at low $z$, if $w_\varphi$ be a dynamic and $z$-dependent value) for $p = 1$. Thus the cosine hyperbolic potential is able to describe both dark matter and quintessence within a tracker framework (Sahni & Wang 2000; Sahni 2004). $\rho_{quint} / \rho_{total} < 0.2$ is necessary in order to satisfy nucleosynthesis constraints. After transition to superfluidity $\rho_{quint}/\rho_B$ ($\rho_B$ of background matter) begins to grow.

As mentioned above, for $p = 1$, $V(\phi) = V_0[\cosh(\lambda\phi) - 1]$ which combine the properties of a tracking solution with a minimum that provides the quintessence at late times; $V(\phi)$ gives a very similar result as one would obtain for a simple exponential potential $V(\phi) \propto e^{-\lambda\phi}$ at early times. Therefore, for $|\lambda\phi| \gg 1$, i.e. far from the minimum it has asymptotic form: $V(\phi) \approx \tilde{V}_0 e^{-\lambda\phi}$; this is similar to exponential quintessence potential: $V(\phi) \approx V_0 e^{-\lambda\phi}$ (Ratra & Peebles 1988; Wetterich 1988a; Ferreira & Joyce 1998) thus the proposed potential preserving some of the desirable properties of the simplest exponential potential case (Peebles & Ratra 2003).

A unified picture of quintessence and SFDM in which both components are described by a pair of scalar fields evolving under the action of the potential but with different values of the exponent $p$; similar values bring together the potential of them:

$$V(\phi,\psi) = V_\phi \left(\cosh(\lambda_\phi \phi) - 1\right)^{p_\phi} + V_\psi \left(\cosh(\lambda_\psi \psi) - 1\right)^{p_\psi} \qquad (6)$$

$p_\psi = 1$ in the case of SFDM and in the case of quintessence the exponent $p_\varphi$ should have a value which the $w_\varphi$ be $\geq 0$. If $p_\varphi = 1$ then there is no difference in the potential of SFDM and quintessence in the present epoch, both of them as the BEC (superfluid) state of boson dark matter. This approach ameliorates the 'coincidence problem' between dark matter (CDM) and quintessence which arises in standard cosmology. It also significantly reduces the discrepancy between the present value of $\rho_m/\rho_r$ and that at the end of inflation (Sahni & Wang 2000).

The collapsing process occurs on the small (ex. galaxy scale) and large scale (clusters and super clusters) in the foam sheets.

Since the strong non-linearity is formed in the very early stage, no extra biasing process is necessary in this model.

2.2 ''Vacuum aether superfluid'' as dark matter and quintessence

*''nature abhors vacuum''*

According to Christianto (2003), we should mention there are at least two groups of approaches in attempting to model perfect fluid as the omnipresent medium in our universe:
(i) the geometrical approach; and (ii) the medium approach.

All theories based on geometrical approach consider vortices as part of their inherent topology structure. In the second model, it is superfluid, which induces vortical structures not the spacetime structure *an sich*. In these groups, the spacetime remains Euclidean, while the superfluid aether drags celestial bodies around. Proponents of this view include:

- Volovik's $^3$He superfluid model
- Winterberg's phonon-roton superfluid vacuum aether
- Consoli's Higgs boson condensate vacuum medium
- Barceló's bosonic induced gravity
- DeAquino's superconducting state model of vacuum quantum

This group sounds like an ad-hoc (inducing aether medium); their models apparently were confirmed by the superfluid-superconducting structure of neutron stars. There are also recent hypothetical boson stars, P-stars, and quark strange stars, which all considers boson in their inherent structures (Christianto 2003).

For applications to cosmology, it is presumed that the "vacuum" is a superfluid-like continuum in which the formation of topological defects as "vortices" generates the stars and galaxies as components of the normal fluid. A question arises as to how to describe the physical origin of wave mechanics of such large-scale structures. This leads to the Winterberg hypothesis (Winterberg 1992, 2002) of the superfluid phonon-roton as a quantum vacuum aether. In effect, the background is the "vacuum aether superfluid" and the stars and galaxies are the "condensed objects" within it (Christianto 2003; Winterberg 2002). In a unified view, universe filled with fluids which unitarily describe as a single vacuum entity either both dark matter and quintessence, so respectively unifying the notion of quintessence and that of all dark components (González-Díaz 2003). An analog of Maxwell's aether as a frictionless fluid with small vortices, vacuum aether superfluid is a quantum fluid made up of Bose and Fermi particles with excitation energy spectrum similar that of helium.

A fluid dynamic model, of course, implies that the vacuum is a medium. As Planck had shown back in 1911, quantum theory demands that the vacuum is not empty but rather filled with the quantum mechanical zero point energy, by Nernst called aether. According to Landau, superfluid

He-II corresponds to a flowing vacuum state, similar to a cosmic aether, in which quasiparticles – bosonic zero modes of quantum vacuum – move. Winterberg (2002) called it 'Planck aether' (aether composed of densely packed Planck mass particles which one may simply call the Planck aether).

Apart from the kind of Bose particle which comprises the vacuum aether superfluid, I investigate the properties of that as the dark components of the universe.

Winterberg proposed a Planck aether which in its ground state is a two component positive-negative mass superfluid with phonon-roton energy spectrum for each component (in the 1970's the discovery of supersymmetry led to the hope that, since bosons and fermions (of identical mass) contribute equally but with opposite sign to the vacuum expectation value of physical quantities, the cosmological constant problem may be resolved by a judicious balance between bosons and fermions in nature. However, it is well known that there is no supersymmetry in our low-energy world. This means that there must by an energy scale $E_{SuSy}$ below which the supersymmetry is violated and thus there is no balance between bosons and fermions in the vacuum energy).

Assuming that the excitation spectrum measured in superfluid helium $^4$He or $^3$He-A is universal. The vacuum of space is a superfluid made up of Bose and Fermi particles, with the particles of the standard model explained as quasiparticle- excitations of this superfluid.

## 3   Superfluidity and universality

According to Landau, superfluid He-II corresponds to a flowing vacuum state, similar to a cosmic aether, in which quasiparticles – bosonic zero modes of quantum vacuum – move. The quanta of these collective modes of the vacuum form a rarefied gas which is responsible for the thermal and viscous effects ascribed to the presence of the normal component. Feynman's first superfluid paper argued that in spite of the inter-atomic interactions, superfluid helium was a Bose-Einstein condensate similar to what happens in the ideal gas where the interactions are absent. In transition to superfluidity, a system of strongly-interacting particles is replaced by a collection of non (or weakly) interacting "elementary excitations" or "quasiparticles". The normal fluid is a gas of excitations where the vacuum is the superfluid. The dense system of strongly interacting $^4$He atoms can be represented in the low-energy corner by a dilute system of weakly interacting quasiparticles (phonons and rotons). In addition, the state without excitation (the ground state, or vacuum) has its own degrees of freedom: it can experience the coherent collective motion. After the works of Onsager (1949) and Feynman (1955), it was understood that rotational degrees of freedom were related to quantized vortices. The lowest energy level related to rotational motion is provided by a vortex ring of minimum possible size whose energy $\sim \hbar^2/ma_0^2$ is in agreement with Landau's suggestion.

Universe fluid in the post-inflationary period behaved as if was a mixture of two different fluids. One of these (boson dark matter) behaves like a superfluid after the transition to superfluidity. The other is a normal viscous fluid (ordinary matter), so above the $T_c$, superfluid transition temperature of boson dark matter, we have a completely viscous universe. The existence of these two flows is seen especially clearly when a rotating superfluid structure rotates about its axis. In this situation the superfluid part remaining at rest. Consequently, the total moment of inertia $I$ of the rotating structure is less than the moment of inertia $I_0$ calculated on the assumption that the whole fluid rotates, and a measurement of the ratio $I/I_0$ enables us to find at once what parts of the fluid are normal and superfluid. The superfluid flow has two other important properties: it does not involve entropy and heat transfer, and it is always potential flow. Hence it follows that a superfluid flow is thermodynamically reversible. Since superfluid flow is potential flow, a steady superfluid flow exerts no force on a solid body (d'Alembert's paradox). The normal flow, on the other hand, exerts a drag force. If the flow is such that the superfluid and normal mass transfers balance, we have a very unusual flow: a force acts on a body immersed in dark matter, but there is no net mass transfer. The *thermo-mechanical effect* in superfluid is as follows: when superfluid flows out of the bulk stream through the thin spaces, a rise in temperature occurs in the bulk fluid. This phenomenon has the natural explanation that the flow of superfluid part transfers no heat, so that the heat remaining in the bulk is distributed over a smaller quantity of matter (Landau & Lifshitz 1987).

The liquid phase of the $^4$He, is currently used as a system for exploring the dynamics of second order phase transitions analogous to the GUT symmetry breaking phase transition of the early universe. The symmetry breaking associated with superconductivity and superfluid transitions is gauge symmetry breaking. The low mass of the helium atom will imply a large value of $E_Z$ (zero-point energy). In the $^4$He, the stable phase at $T = 0$; $P = 0$ must be the *liquid*.

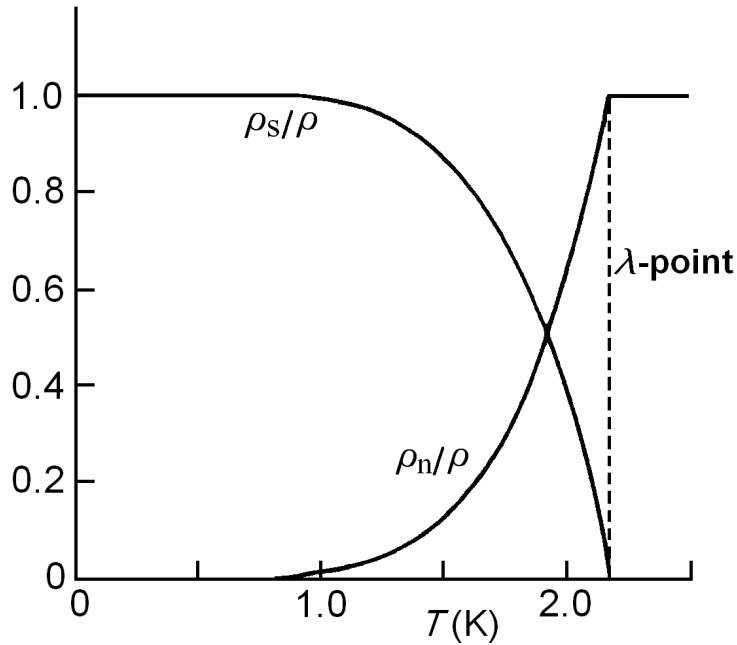

Figure 1. Normal and superfluid densities of HeII, $\rho_n$ and $\rho_s$, divided by the total density $\rho$ of the liquid, as functions of temperature T; (Andronikashvili 1971).

3.1 superfluidity and wave-particle duality

The two-fluid model of superfluid is a non-fundamental phenomenological explanation presented by Landau (1941). We assume that we are dealing with HeII in the linear regime: i.e. all velocities, temperature gradients, pressure gradients and so on will be assumed to be not too large. These ideas can be developed quantitatively by deriving the relevant equations of motion for the two-fluid system. They are sometimes known as the thermohydrodynamical equations because of the peculiar necessity of intermingling thermodynamics and hydrodynamics in their derivation. Based on Landau theory, two fluids interpenetrate freely, passing through each other entirely without interaction but where velocities, or gradients in $T$ and $P$, become sufficiently large, quantized vortices are created in the superfluid component. These result in the development of a so-called mutual friction between the two components whenever their relative velocity is non-zero, in partial violation of this premise of the two-fluid model. In most physical situations, the measured critical velocities in HeII are found to be much smaller than $v_{\text{Landau}}$. This is because of the assumption, that excitation spectrum for helium II includes all the permitted low-lying excited states of the liquid, is not entirely justified. While it is true that the indicated dispersion curve includes all permitted elementary excitations, it turns out that other (metastable) excitations involving quantized vortex rings and lines can also exist and that their creation is often characterized by critical velocities which are smaller than $v_{\text{Landau}}$. Also the Landau critical velocity $v_{\text{Landau}}$ is vanishingly small in $^3$He-A because of the gapless fermionic energy spectrum ($v_{\text{Landau}} \sim \Delta/p_0$). Also derived from Landau theory the superfluid component carries no entropy and experiences no flow resistance whatever; that is, its viscosity is identically zero and no turbulence can be created in it, $\nabla \times v_s = 0$ so the superfluid flow is potential, but later Onsager (1949) and Feynman (1955) found that this statement must be generalized: $\nabla \times v_s \neq 0$ at singular lines, the quantized vortices, around which the phase of the order parameter winds by $2\pi N$. The discovery of superfluid $^3$He-A further weakened the rule: the nonsingular vorticity can be produced by the regular texture of the order parameter – the wave function of Cooper pair with spin $S = 1$ and orbital momentum $L = 1$ – according to the Mermin-Ho relation (Mermin & Ho 1976). As distinct from superfluid $^4$He, in $^3$He-A there is no energy gap between rotational and irrotational motions. This is the reason why the bulk $^3$He-A behaves as normal fluid. Thus none of the conditions representing the signatures of superfluidity in Landau's sense are strictly valid in a system like $^3$He-A (Volovik 2003).

Superfluidity must be explained as macroscopic quantum phenomena. The existence of Bose-Einstein condensation does not directly imply superfluidity, because an ideal Bose gas is not a superfluid. The lambda transition does, in fact, correspond to Bose-Einstein condensation, albeit in a modified form because of the presence of interatomic forces. The superfluid vacuum is thus more complicated than its predecessor – the Bose–Einstein condensate. For example, experiments show that the condensate fraction of superfluid $^4$He at $T = 0$ is c. 14%, rather than the 100% expected of an ideal gas (see Fig. 2) and this is an expected consequence of the finite interatomic forces, which cause some of the particles to occupy what are called depletion levels, even at $T = 0$, depleting the fraction of particles in the condensate, although, the trends of the increment in superfluid and condensate fraction is rather identical. Nevertheless, like some cosmological literature (Nishiyama, Morita & Morikawa 2004; Silverman & Mallett 2002), sometimes in this theory I use Bose–Einstein condensation instead of transition to superfluidity.

Also, in spite of the knowledge that Landau two-fluid model of superfluidity is a non-fundamental and defective theory, in some parts of this theory, we are compelled to utilize it.

Although the two-fluid model describes most aspects of the behavior of HeII remarkably well, it is important always to remember that there can be no suggestion of two physically distinct materials being present. HeII is an assembly of $^4$He atoms, all of which are identical. The rotating cylinders measure the normal fluid; the thin tubes measure the superfluid flow. One of these motions is normal, has the same properties as the motion of an ordinary viscous fluid, but the other is the motion of superfluid. Maybe the two motions occur without any transfer of momentum (Landau & Lifshitz 1987).

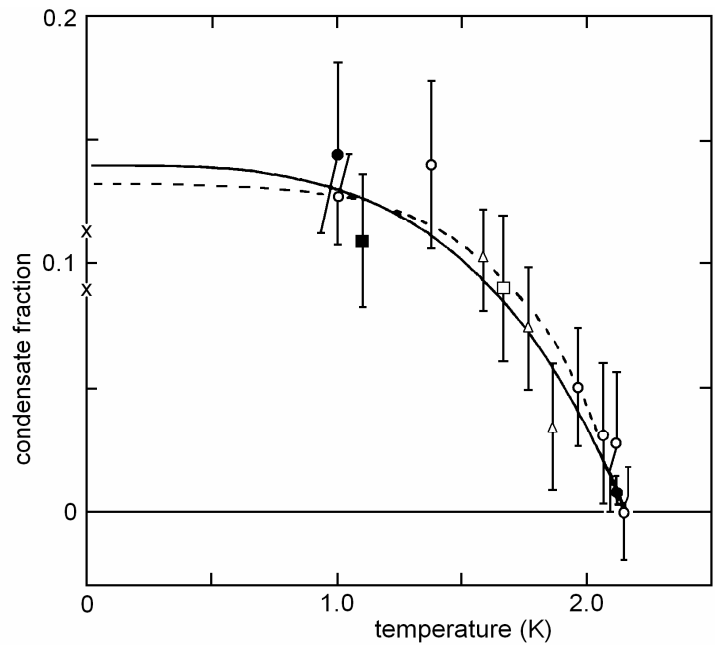

Figure 2. The condensate fraction of liquid $^4$He (Sears et al. 1982; Campbell 1983). The full curve is a theoretical temperature-dependence fitted to the data points.

"Light" like the "superfluid dark matter" has a dual behavior. Surprisingly, we cannot separate these two behaviors in each case rather we can only distinguish them. "*Wave* and *particle* behavior of light" and "*superfluid* and *normal* flow of superfluid dark matter" are coupled with each other. Superfluid dark matter is reminiscent of the aether and modeling the universe using superfluid aether is compatible.

(Chiao 2000) showed the possibility that photons, which are bosons, can form a 2D superfluid due to Bose-Einstein condensation inside a nonlinear Fabry-Perot cavity filled with atoms in their ground states. The effective mass and chemical potential for a photon inside this fluid are nonvanishing. This implies the existence of a Bogoliubov dispersion relation for the low-lying elementary excitations of the photon fluid, and in particular, that sound waves exist for long-wavelength, low-frequency disturbances of this fluid. These fascinating experiments provide evidence for the superfluid nature of photons.

The world consists of interacting particles and waves. In our ordinary perception of reality, we are fairly confident that we can distinguish easily between waves and particles, but in fact they can sometimes switch identities. Light waves may be regarded as a stream of massless particles, also called photons, and particles of matter sometimes have a wavelike nature. Generally speaking, the corresponding wavelength is extremely short, but for atoms that move slowly, it may be observable. Einstein predicted that if a gas is cooled to very low temperatures, all the atoms should gather in the lowest energy state. Matter waves of the individual atoms then merge into a single wave; indeed, they can be said to "sing in unison." Thousands of atoms behave like one big super atom. This is Bose-Einstein condensation (Sune Svanberg, Nobel Prize in Physics 2001).

Laser light as a Bose-Einstein condensate

It should be noted that the most familiar examples of the situation where particles occupy a single wave state, so that the wave function can be considered classical, are electromagnetic waves and laser lights. The invention of the laser led to the discovery of a new state of light, namely the coherent state, which is a very robust one. Ketterle was also able to extract a beam of coherent matter from the condensate, thus achieving the first atom laser. An ordinary laser yields coherent radiation, an atom laser a stream of coherent matter.

Bose-Einstein condensates are formed by cooling gas atoms to a fraction of a degree above absolute zero. At that temperature, the atoms all drop into the same quantum state. That makes them coherent, all possessing the same quantum wave function, a state comparable to that of photons in laser systems. Cubic nonlinearities show up in the Landau-Ginzburg equations for coherent macroscopic order parameters or giant effective quantum wavefunctions in second order phase transitions including superfluids and lasers.

It is now believed that most systems displaying free motion of particles well below some characteristic degeneracy temperature transform to the superfluid state or undergo condensation.

According to the de Broglie theory of wave-particle duality of everything, this duality can be interpreted as superfluidity of everything in the universe but based on the extended and unified model of superfluidity as described above. The Universe includes fermionic quasiparticles in Fermi superfluids and bosonic quasiparticles in Bose superfluids.

Bose-Einstein condensate behaves as a single quantum object despite its size. Because of this behaviour, it is hoped that fundamental tests of Quantum Mechanics can be made. An example of this Quantum Mechanical behaviour is the interference between two condensates. Just as water ripples or light waves interfere (as in Young's double slit experiment), two separate condensates can be released from their traps and interfere as they overlap. This was first observed in 1996 at MIT, demonstrating the wave-like behaviour of the condensate.

Every fermionic material must pair (BCS or Cooper pair, etc.) before transition to superfluidity or Bose-Einstein condensation. The basic concept of the BCS-BEC crossover is as follows: As long as the attractive interaction between fermions is weak, the system exhibits the superfluidity characterized by the energy gap in the BCS mechanism. On the other hand, if the attractive interaction is strong enough, the fermions first form bound molecules (bosons). Then they start to condense into the bosonic zero-mode at some critical temperature. These two situations are smoothly connected without the phase transition. One of the most striking features of the crossover is that the critical temperature in the BEC region is independent of the coupling for the attraction between fermions. This is because the increase of the coupling only affects the internal structure of the bosons, while the critical temperature is determined by the boson's kinetic energy.

The naive calculations of the energy density of the superfluid ground state in the framework of the effective theory suggests that it is on the order of the zero-point energy of the bosonic field in Bose superfluids and on the order of the energy of the Dirac vacuum of fermionic quasiparticles in Fermi superfluids.

Our Universe contains quantum Bose and Fermi fields, describing the interaction of the elementary quasiparticles (fermionic quasiparticles and collective bosons) with each other and with the quantum [bosonic and fermionic] vacuums. In this model of wholly superfluid universe, leptons, gauge fields, Higgs fields, gravitons, W-bosons and etc. are interacting Fermi and Bose quantum fields.

An effective theory in condensed matter does not depend on details of microscopic (atomic) structure of the substance. Once the universality class is determined, the low-energy properties of the condensed matter system are completely described by the effective theory, and the information on the underlying microscopic physics is lost. The microscopic details provide us only with the "fundamental constants", which enter the effective phenomenological Lagrangian, such as speed of "light" (say, the speed of sound), superfluid density, modulus of elasticity, etc.

Volovik (2003) described that various cosmological phenomena could be represented using $^3$He superfluid medium. He argued that the Universe and its broken-symmetry ground state (the physical vacuum) could be represented with condensed matter system.

Ordinary baryonic matter acts as a part of fermionic quasiparticles of quantum fermionic vacuum or impurities in the superfluid boson dark matter. If accumulation of ordinary baryonic matter on the interface of two phases (surface of hydrodynamic bubbles or other interfaces) be energetically favorable (like the formation of electron bubble; when an electron as a fermion is injected into bosonic helium, it forms a spherical cavity- an "electron bubble"- from which helium atoms are excluded. The bubble forms because the electron is strongly repelled by helium) therefore it stabilizes the bubbles (voids) and cosmic foam on the interface of two phases. Also, magnetic field can affect the stability of MHD bubbles. In this study we must pay attention to the effect of adsorbed ordinary baryonic matter on the surface of bosonic sea of dark matter and bubbles, on the motion of bubbles and hydrodynamical properties of the interface.

In the course of the expansion of universe, at high temperature the bosonic universe fluid has all the symmetries that ordinary condensed matter can have: translational invariance, global $U(1)$ group and... . When the temperature decreases further the boson dark matter reaches the superfluid transition temperature $T_c$, below which it spontaneously loses each of its symmetries except for the translational one - it is still liquid.

At low energy corner, the fermionic quasiparticles in the vicinity of the Fermi point - the fermion zero modes - represent chiral (left-handed and right-handed) fermions; the collective bosonic modes of superfluid boson dark matter represent some gauge and gravitational fields acting on the chiral quasiparticles. They all gradually emerge and arise as fermionic and bosonic collective modes of such a system. Also, this emerges together with the 'laws of physics' and together with such notions as chirality, spin, isotopic spin, etc. At high energy (on the Planck scale) these modes merge with the continuum of all the high-energy degrees of freedom of the 'Planck condensed matter' (Volovik 2003) and thus can no longer be separated from each other.

On the basis of Landau two-fluid model of superfluidity, the fermionic quasiparticles and collective bosons (normal component of boson dark matter) perceive the homogeneous ground states of condensed bosonic and fermionic matters as an empty space - a vacuum - since they do not scatter on particles comprising this vacuum state: quasiparticles move in a quantum liquid without friction just as particles move in empty space. For the perception of quasiparticles, the value of Landau critical velocity $v_{Landau}$ is determinant, although, on the basis of the extended and unified model of superfluidity instead of Landau two distinct-fluid model, this interpretation must be slightly modified.

In my model, inspired by low-temperature $^4$He, we argue that dark matter content of the universe can undergo Bose condensation. At the beginning of the dark ages after transition to superfluidity and electromagnetically induction of gravity, the superfluid boson dark matter interacts strongly with ordinary baryonic matter, as well as with itself, so instead of flying past each other, as in a gas, the whole liquid moves more coherently. For $T < T_c$, the dark matter will be superfluid and according to Landau two-fluid model, *Superfluid flow/part* of superfluid dark matter is always potential flow and it does not involve entropy and heat transfer, a steady superfluid flow exerts no force on a body and the *normal flow/part*, on the other hand, exerts a drag force. Along with aging of the universe (further cooling), the superfluid part (in BEC state) of superfluid dark matter will increase. Therefore the superfluid boson dark matter seems to gradually decouple from ordinary baryonic matter after the transition to superfluidity and BEC state. It is not surprising therefore that there is an increasing amount of investigation in using superfluid model to represent cosmology dynamics (Zurek 1996; Volovik 2001, 2002, 2003; Silverman & Mallett 2002).

Matter and anti-matter are created in equal amounts in the universe, however, in cosmology, for instance, people are trying to explain why there is more matter than anti-matter in the universe. Existence of bare particles comprising the quantum vacuum of quantum liquids represents the main difference from the relativistic quantum field theory (RQFT). In RQFT, particles and antiparticles which can be created from the quantum vacuum are similar to quasiparticles in quantum liquids. According to Volovik (2003) [based on fermions], positive energy states correspond to real particles, the negative energy states to the Dirac Sea, which is the (quantum) vacuum; a hole in the Dirac Sea corresponds to an antiparticle, with effectively a positive energy. The creation of a real particle usually takes place when a particle in the Dirac sea of negative energy states is excited to a positive energy state. Then matter and anti-matter are created in equal amounts because of the simultaneous creation of a hole, i.e. an antiparticle. The process leading to matter creation can be studied and will be explained in the context of superfluid helium.

3.1 Gravity

Bose-Einstein condensates (BEC) have recently been the subject of considerable study as possible analogue models of general relativity (Garay *et al.* 2000; Garay *et al.* 2001; Barceló *et al.* 2001; Barceló *et al.* 2003; Novello *et al* 2002; Ferrer & Grifols 2001). The basic idea that gravity is a semi-classical, long-wavelength effect induced by an underlying quantum field theory is now more than twenty years old. This approach is very appealing since, in fact, one can get a picture of the world with only three elementary interactions and where the origin of the fourth, gravitation, has to be searched in the structure of the vacuum. This leads to consider the implications of spontaneous symmetry breaking through an elementary scalar field. Their picture suggests that Einstein general relativity may represent the weak field approximation of a theory generated from flat space with a sequence of conformal transformations. This explains naturally the absence of a large cosmological constant from symmetry breaking (Consoli 2000). The central idea of a condensed matter analog model of GR is that the modifications to the propagation of a field/wave due to curved spacetime can be reproduced (at least partially) by an analog field/wave propagating in some material background with space and time dependent properties (Barceló *et al.* 2001). As Garay *et al.* (2000, 2001) have shown, perturbations in the phase of the condensate wavefunction satisfy, in the low-momentum regime, an equation equivalent to that of a massless scalar field in a curved spacetime (the d'Alembertian equation $\Delta \phi = 0$), but with the spacetime metric replaced by an effective metric that depends on the characteristics of the background condensate.

The condensed matter analogy gives us examples of how the effective gravity appears as an emergent phenomenon in the low-energy corner. Quantum liquids also provide examples of how the metric field naturally emerges as the low-energy collective mode of the quantum vacuum. The action for this mode is provided by the dynamics of quantum vacuum in accordance with Sakharov's theory, and even the curvature term can be reproduced in some condensed matter systems in some limiting cases. From this point of view, gravity is not the fundamental force, but is determined by the properties of the quantum vacuum: gravity is one of the collective modes of the quantum bosonic vacuum.

If gravity emerges in the low-energy corner as a low-energy soft mode (zero mode) of the underlying quantum boson dark matter, then it would indicate that quantum gravity simply does not exist. If there are low-energy modes which can be identified with gravity, it does not mean that these modes will survive at high energy. Most probably they will merge with the continuum of all other high-energy degrees of freedom of the Planck condensed matter (corresponding to the motion of separate atoms of the liquid in the case of $^4$He and $^3$He) and thus can no longer be identified as gravitational modes. The condensed matter analogy supports the extreme point of view expressed by Hu (1996) that one should not quantize gravity again. What is allowed in effective theory is to quantize the low-energy modes to produce phonons from sound waves and gravitons from gravitational waves but one should not use the low-energy quantization for the construction of Feynman loop diagrams containing integration over high momenta. Thus deeper quantum theory of gravity makes no sense in this philosophy. Our knowledge of the physics of phonons/gravitons does not allow us to make predictions on the microscopic structure of the bosonic vacuum (Volovik 2000, 2003).

The vacuum energy cannot interact with gravity, and cannot serve as a source of the effective gravity field: the vacuum is not gravitating. On the other hand, the long-wavelength perturbations of the vacuum are within the sphere of influence of the low-energy effective theory, and such perturbations can be the source of the effective gravitational field. Deviations of the vacuum from its equilibrium state, described by the effective theory, are gravitating.

Consoli (2000, 2002) described gravitation was originated from Higgs boson condensate. Consoli (2000) argued that the basic idea that gravity can be a long-wavelength effect induced by the peculiar ground state of an underlying quantum field theory leads to consider the implications of spontaneous symmetry breaking through an elementary scalar field. He point out that Bose-Einstein condensation implies the existence of long-range order and of a gap-less mode of the Higgs-field. This gives rise to a 1/r potential and couples with infinitesimal strength to the inertial mass of known particles. If this is interpreted as the origin of Newtonian gravity one finds a natural solution of the hierarchy problem. Consoli (2000, 2002) has also considered similarity between his condensate model and superfluid aether hypothesis, in the spirit of Landau. It is obvious that in a description where gravity is a long-wavelength excitation of the scalar condensate there are differences with respect to the standard ideas. For instance, the gravitational force is naturally instantaneous.

Electromagnetically Induced ''Gravity''

Landau and Khalatnikov (see the book by Khalatnikov (1965)) proposed the simplest effective field theory of superfluids, where only the gravitational field appears as an effective field. In this theory, a weakly excited state of the collection of interacting $^4$He atoms can be considered as a small number of elementary excitations – quasiparticles, phonons and rotons. In addition, the state without excitation – the ground state or vacuum – can have collective degrees of freedom. The superfluid vacuum can move without friction, and inhomogeneity of the flow serves as the gravitational and/or other effective fields. The matter propagating in the presence of this background is represented by fermionic (in Fermi superfluids) or bosonic (in Bose superfluids) quasiparticles, which form the so called normal component of the liquid. Such two fluid hydrodynamics introduced by Landau and Khalatnikov (Khalatnikov 1965) is the example of the effective field theory which incorporates the motion of both the superfluid background (gravitational field) and excitations (matter). This is the counterpart of the Einstein equations, which incorporate both gravity and matter. One must distinguish between the particles and quasiparticles in superfluids. The particles describes the system on a microscopic level, these are atoms of the underlying liquid ($^3$He or $^4$He atoms). The many-body system of the interacting atoms forms the quantum vacuum – the ground state. The conservation laws experienced by the atoms and their quantum coherence in the superfluid state determine the low frequency dynamics – the hydrodynamics – of the collective variables of the superfluid vacuum. The quasiparticles – fermionic and bosonic – are the low energy excitations above the vacuum state. They form the normal component of the liquid which determines the thermal and kinetic low-energy properties of the liquid. Note that for the curl-free superfluids the sound waves represent the only "gravitational" degree of freedom. The Lagrangian for these "gravitational waves" propagating above the smoothly varying background is obtained by decomposition of the superfluid velocity and density into the smooth and fluctuating parts: $v_s = v_{s\ smooth} + \nabla \alpha$.

The energy spectrum of sound wave quanta, phonons, which represent the "gravitons" in this effective gravity, is determined by

$$g^{\mu\nu} p_\mu p_\nu = 0, \text{ or } (\widetilde{E} - \mathbf{p}.\mathbf{v_s})^2 = c^2 p^2. \tag{7}$$

In Bose superfluids the fermionic degrees of freedom are absent, that is why the quantum field theory there is too restrictive, but nevertheless it is useful to consider it since it provides the simplest example of the effective theory. On the other hand the Landau-Khalatnikov scheme is rather universal and is easily extended to superfluids with more complicated order parameter and with fermionic degrees of freedom. We can expect that the analysis of the condensed matter analogues of the effective gravity, in particular, of the Landau-Khalatnikov two-fluid hydrodynamics and its extensions will allow us to solve the longstanding problem of the cosmological constant (Volovik 2000).

O'Dell *et al.* (1999) in an exciting study showed that particular configurations of intense off- resonant laser beams (simulate the Bose-Einstein condensate) can give rise to an attractive 1/r interatomic potential between atoms located well within the laser wavelength. Such an $r^{-1}$ attractive potential can simulate gravity between quantum particles. Such a "gravitational-like" interaction is shown to give stable Bose condensates that are self-bound (without an additional trap) with unique scaling properties and measurably distinct signatures. This suggests it might be possible to study gravitational effects, normally only important on the stellar scale, in the laboratory. Particularly interesting is the possibility of experimentally emulating Boson stars (a system of gravitating Bosons): gravitationally bound condensed Boson configurations of finite volume, in which the zero-point kinetic energy balances the gravitational attraction and thus stabilizes the system against collapse.

Ferrer & Grifols (2004) showed that scalar mediated interactions among baryons extend well above the Compton wavelength, when they are embedded in a Bose-Einstein condensate composed of the mediating particles. Indeed, this non-trivial environment results in an infinite-ranged interaction. They have entertained the possibility that the dark matter is composed by a Bose-Einstein condensate of light scalar particles coupled to the baryonic component of the Universe. The long-range forces among baryons caused by the pair exchange of dark matter particles in the presence of the condensate. It has been recently proposed that the observed 511 keV emission from the Galactic bulge could be the product of very light annihilating scalar dark matter particles. Inspection of these results immediately leads us to realize an important consequence of Bose-Einstein condensation. Namely, at low temperature (i.e., below $T_c$), the force, that was finite ranged at high temperature (i.e., above $T_c$), becomes infinite ranged.

The instantaneous nature of the long-range gravitational interaction is a direct consequence of its non-local origin from the scalar condensate. Indeed, one can imagine that removing at spatial infinity all gravitational sources produces an infinite flow of the scalar condensate and, as a net result, the vacuum becomes 'empty' around a given body and its inertia vanishes. Therefore, one cannot speak of absolute accelerations with respect to empty space since the inertial mass of a test particle depends on the existence of the scalar condensate whose density is determined by the distribution of gravitating matter (Consoli 2000). In this sense, the 'Mach Principle' represents a concise formulation of the inextricable connection between inertia and gravity due to their common origin from the same physical phenomenon: the condensation of the [bosonic] scalar field.

3.2 Black hole

Though the corresponding 'Einstein equations' for 'gravity' itself are not covariant, by using the proper superflow fields we can simulate many phenomena related to the classical and quantum behavior of matter in curved space-time, including black-hole physics (Volovik 2000).

There is a close analogy between sound propagation on a background hydrodynamic flow, and field propagation in a curved spacetime; and although hydrodynamics is only a long-wavelength effective theory for physical (super)fluids, so also field theory in curved spacetime is to be considered a long-wavelength approximation to quantum gravity (Garay et al. 2000).

The physical mechanism of the sonic black hole is quite simple: inside the horizon, the background flow speed (speed of the condensate flow) $v$ is larger than the local speed of sound $c$, and so sound waves are dragged inwards and these regions very closely mimic the key kinematic features of black hole physics (Garay et al. 2000; Barceló *et al.* 2001). (Unruh 1981, 1995) urged a specific motivation for examining the hydrodynamic analogue of an event horizon, namely that as an experimentally and theoretically accessible phenomenon it might shed some light on the Hawking effect (thermal radiation from black holes, stationary insofar as the back reaction is negligible).

Condensates can be produced in a vortex state, wherein the gas rotates, reminiscent of water going down the drain. A pulse of slow light traveling through a vortex would find itself dragged along with the gas- very similar to a phenomenon expected to occur near black holes. With slow light, we can study this and some other black hole phenomena in the laboratory.

3.3 Quark-gluon plasma

Dense quark matter at low temperatures is expected to be in a BCS-paired (Bardeen-Cooper-Schrieffer pairing before they can collapse into a Bose condensate or superfluid) superfluid state induced by attractive coupling in the color-$\bar{3}$ channel, with the pairs having zero total angular momentum, $J$ (Iida & Baym 2002). Therefore the essence of gluon like the graviton which is the superfluid vacuum (condensate) of superfluid bosonic dark matter or 'light', is the superfluid vacuum of superfluid fermionic quark matter and the coherent and strongly coupled quark-gluon plasma is actually the BCS-paired superfluid quark matter. Recently, physicists at MIT have found conclusive evidence for superfluidity in an ultracold Fermi gas by observing quantized vortices in a rotating gas of $^6$Li atoms. The results could shed new light on systems as diverse as high-temperature superconductors, neutron stars and the quark-gluon plasma. This is the first ''high-temperature'' superfluid, measured in terms of the ratio of the critical temperature ($T_c$), at which the superfluid transition takes place, to the Fermi temperature ($T_f$).

Strongly Coupled Quark Gluon Plasma (SCQGP) is a new state of matter which should have existed a few microseconds after the Big Bang. Recently, researchers at the Relativistic Heavy Ion Collider (RHIC) at Brookhaven National Laboratory recreate the quark-gluon plasma (QGP). Their experiments suggest the quarks in the plasma are exhibiting incredibly synchronized group behavior and interacting strongly with each other and the surrounding gluons. This makes the plasma more similar to a liquid than a gas. In fact, the strength of interactions in the quark-gluon plasma makes it the most ideal liquid ever observed- 10 to 20 times as liquid-like as water. It seems that from the time of the big bang until 10 μsec later, the universe was liquid: "*And the assumption is that provided you can create a large enough droplet, it will evolve in the same way as the early Universe.*"

Superfluid quark matter is an ultrarelativistic color superconductor that is homogeneous, in the sense that the real gluon field vanishes everywhere and the order parameter is everywhere continuous in magnitude and orientation (Iida & Baym 2001). Such homogeneity is similar to that in superfluid $^3$He and superfluid neutron matter, as noted by (Bailin & Love 1984), because in both cases breaking of the global $U$(1) gauge symmetry is accompanied by global symmetry breaking associated with the internal degrees of freedom. In superfluid quark matter, the possible order parameters are generally *anisotropic* in color space, a situation analogous to superfluid $^3$He in which, as seen experimentally, the anisotropy lies in spin space.

Color superconductivity in quark matter becomes an astrophysically interesting problem if neutron star interiors are sufficiently dense that they contain quark matter cores. Generally, a quark superfluid in a neutron star would not be electrically neutral since each quark has fractional electric charge; rather it would coexist with electrons (and muons) in such a way as to ensure electric neutrality in the system. Because of the dually charged nature of the quarks, macroscopic manifestations of both color and electromagnetic superconductivity such as Meissner effects, generation of London fields, and vortex formation are expected from magnetic fields and rotations as observed in these celestial objects. Iida & Baym (2001) discussed these issues, which may be relevant to magnetic structure, cooling, and rotational evolution of the neutron stars.

# 4 Time varying dark components result in time varying constants

## 4.1 time varying constants

Hypotheses about the time varying electromagnetic fine structure constant (α) which are proposed by (Aspden & Eagles 1972) and (Winterberg 1992, 2002) are deficient and even in some parts incorrect, although the overall picture of the vacuum aether constituents is similar to our theory (in the theory if we attribute opposite signs to the bosonic and fermionic superfluids, this representation is analogous to the SuSy and above authors' thoughts), therefore, inspired by (Aspden & Eagles 1972; Winterberg 1992, 2002) and according to Mach's Principle, the electromagnetic fine structure constant (α) depends on the characteristic and dynamic properties of the dark components (aether) such as quintessence energy density of the universe. Along with aging of the universe, increment in the quintessence (condensate) fraction of the universe with a trend similar to that of superfluid $^4$He (decelerated rate of increase in the quintessence energy density) result in a monotonically increment in α with alike trend. Surprisingly, this picture is fully consistent with cosmological observations (Fig. 3).

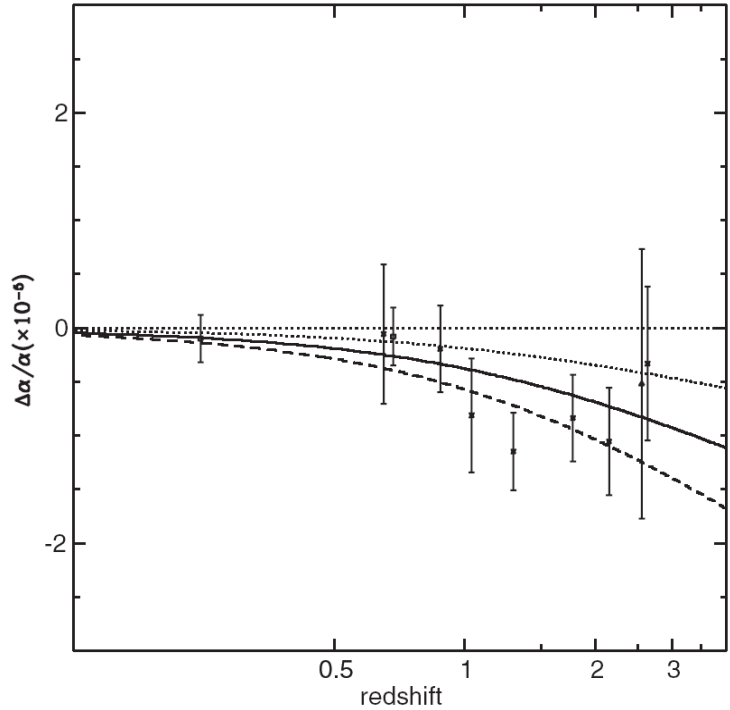

Figure 3. The data points are the QSO results for the changing *α* (Magueijo 2003). The solid line depicts theoretical prediction in several varying-*α* models.

As mentioned in previous section, studies have attempted to determine whether quintessential scalar fields could explain both cosmological dark matter and the recent acceleration of the universe. The most observationally sensitive 'constant' is the electromagnetic fine structure constant, $α = e^2/\hbar c$ and recent observations motivate the formulation of varying-α theories. (Shea 2005) has made a list of the criterions measuring change in the fine structure constant:

1- Quasars
2- Cosmic Microwave Background
3- Big bang nucleosynthesis
4- The Oklo Reactor
5- Rhenium Isotopes
6- Atomic Clocks

Quasar (QSO) spectra

The new observational many-multiplet technique of Webb *et al.* (1999), exploits the extra sensitivity gained by studying relativistic transitions to different ground states using absorption lines in quasar (QSO) spectra at medium redshift. It has provided the first evidence that the fine structure constant might change with cosmological time. The trend of these results is that the value of α was lower in the past, with $\Delta\alpha/\alpha = -0.72\pm0.18\times10^{-5}$ for $z \approx 0.5-3.5$. The ''many-multiplet'' method additionally improves on the calculations in a second respect in that it minimizes possible systematic sources of experimental error. The only identified source for a systematic error, atmospheric dispersion and isotopic abundance evolution, would only accentuate the change in α.

Cosmic Microwave Background and Big Bang Nucleosynthesis

Recent measurements of a peak in the angular power spectrum of the cosmic microwave background appear to suggest that geometry of the universe is close to being flat. But if other accepted indicators of cosmological parameters are also correct then the best fit model is marginally closed, with the peak in the spectrum at larger scales than in a flat universe. Such observations can be reconciled with a flat universe if the fine structure constant had a lower value at earlier times (Battye, Crittenden & Weller 2001), which would delay the recombination of electrons and protons and also act to suppress secondary oscillations as observed. Battye, Crittenden & Weller (2001) discussed evidence for a few percent increase in the fine structure constant between the time of recombination and the present.

A smaller value of α in the past would change the ionization history of the universe, postponing the recombination of electrons and protons, i.e. last-scattering would occur at lower redshift. It would also alter the ratio of baryons to photons at last-scattering, leading to changes in both the amplitudes and positions of features in the CMB power spectrum, primarily at angular scales ≤ 1°, therefore, limits on the temperature fluctuations of the CMB could lead to a measurement with experimental sensitivity of $|\Delta\alpha/\alpha| < 10^{-2} - 10^{-3}$, at z ~ 1000 (Hannestad 1999;

Kaplinghat, Scherrer & Turner 1999; Avelino, Martins & Rocha 2000; Battye, Crittenden & Weller 2001; Kujat & Scherrer 2000). Analysis of data from the Wilkinson Microwave Anisotropy Probe (WMAP) provides a bound −0.06 < Δα/α < 0.02 at 95% CL. More refined measurements of the cosmic microwave background could feasibly check for this and would lead to an independent verification of the quasar results.

(Battye, Crittenden & Weller 2001) mentioned that realistic models in which α vary may contain one or more light scalar fields which mediate the precise variation. Clearly, if such a field exists it should be included in the calculation of the CMB anisotropies in the Boltzmann hierarchy of CMBFAST, either explicitly or as an extra relativistic degree of freedom. This would allow a subsequent analysis to include effects of the time variation of α, rather than just a change between the time of recombination and the present day. It may even be possible for such a field to act as a quintessence field, removing the need for Λ.

Other investigations have claimed preferred non-zero values of Δα < 0 to best fit the Big Bang Nucleosynthesis (BBN) data at z ≈ $10^{10}$ (Avelino *et al.* 2000; Battye, Crittenden & Weller 2001).

Barrow & Mota (2003) calculated that α may behave differently within the galaxy than inside emptier regions of space. Once a young galaxy condenses and relaxes into gravitational equilibrium, α nearly stops changing inside it but keeps on changing outside. In contrast to current belief, in my model, quintessence can be a fluid or quantum liquid as a consequence like Bekenstein-type models (where changes in α are driven by the matter density) my model allows local spatial variation, which could provide different dynamics for Δα in our local environment (Oklo and meteorites) and over cosmological scales (QSO data). Thus, the terrestrial experiments that probe the constancy of α suffer from a selection bias.

It is found that at present

$$\frac{\dot{\alpha}}{\alpha} \approx 2.98 \times 10^{-16} \, h \, \text{year}^{-1}$$

with $H_0 = 100 \, h$ km s$^{-1}$ Mpc$^{-1}$, $\Omega_\Lambda = 0.71$ and $\Omega_m = 0.29$. For $h = 0.7$ this gives a fractional variation in alpha of about $2 \times 10^{-16}$ per year, which should soon be within the reach of technology. Such an observation would be an incredible vindication of the Webb results. On the other hand, this effect would become a further annoyance for those concerned with the practicalities of defining the unit of time. Without lambda the current rate of variation in *α* would be of the order of $10^{-14}$ per year.

Another remarkable set of recent observations is of Type Ia supernovae in distant galaxies. These data have extended the Hubble diagram to redshifts, $z \geq 1$. They imply an accelerated expansion of the universe. When combined with CMB data, the supernovae observations favour a flat universe with approximate matter density, $\Omega_m \approx 0.3$ and vacuum energy density, $\Omega_{quint} \approx 0.7$. The idea that the fine structure constant is varying over the history of the universe has a long history, and if this discovery is confirmed by further observations and analyses, then it will have a profound impact on the future of physics.

Theories have been considered which introduce new scalar fields whose couplings with the Maxwell scalar $F_{\mu\nu}F^{\mu\nu}$ allow time varying α. There are a variety of possible physical expressions of a changing α. Bekenstein proposed a varying *e* theory (Bekenstein 1982). Sandvik, Barrow & Magueijo (2002) show that by applying a generalisation of Bekenstein's varying-*e* theory in a cosmological setting including the cosmological constant, Λ, we are able to explain the magnitude and sense of the observed change in α. The main assumption is that the cold dark matter has magnetic fields dominating their electric fields. The magnetostatic energy then drives changes in α in the matter dominated epoch, but as the Universe starts to accelerate these changes become friction dominated and come to a halt. This gives a decelerated rate of change in α, just as the universe starts to accelerate, in accord with both data sets. The only energy scale they introduce is of the order of Planck scale, which also makes the model attractive. Sandvik, Barrow & Magueijo (2002) have shown how a cosmological generalization (including the cosmological constant, Λ) of Bekenstein's theory of a varying *e* (the electromagnetic coupling) can naturally explain the reported variations in the fine structure constant. Bekenstein's original theory takes *c* and $\hbar$ to be constants and attributes variations in α to changes in *e*, or the permittivity of free space. This is done by letting *e* take on the value of a real scalar field which varies in space and time $e_0 \rightarrow e = e_0 \epsilon (x^\mu)$, where $\epsilon$ is a dimensionless scalar field and $e_0$ is a constant denoting the present value of *e*. This theory generalizes Bekenstein's approach by including the effects of the varying $\epsilon$ (or $\psi$) field on the gravitational dynamics of the expanding universe. The scalar field $\psi$ couples only to electromagnetic energy. They find that in this theory the $\psi$ and α, remains almost constant in the radiation era where baryonic species become relativistic, undergoes small increase in the matter era, but approaches a constant value when the universe starts accelerating because of the presence of a positive cosmological constant. This trend in $\psi$ and α is compatible with my unified theory of time varying boson scalar dark components.

The model of Sandvik, Barrow & Magueijo (2002) can be seen as a more conservative alternative to (Moffat 1993; Barrow 1999; Albrecht &

Magueijo 1999; Barrow & Magueijo 2000; Moffat 2001; Magueijo 2003), where a VSL (varying speed of light) scenario was proposed which could explain the observed acceleration of the universe and variations in α, as well as their remarkable coincidence in redshift space. Barrow & Magueijo (2000) proposed that the Supernovae results, implying evidence for an accelerating Universe may be closely related to the recent discovery of redshift dependence in the fine structure constant α They examined a class of varying-$c$ theories in which changes in $c$ are driven by a scalar field which is coupled to the gravitational effect of pressure. (Moffat 2001) like (Barrow & Magueijo 2000) has studied a simple model that incorporates a VSL behavior in a cosmological setting but assuming that the electric charge $e$ and Planck's constant $\hbar$ are truly constants of nature. They have shown that a varying speed of light can explain the reported variations in the fine structure constant, while satisfying all the observational bounds (fits the spectral line data). (Moffat 2003a) proposed a model in VSGW metric frame in which, the speed of light is constant but the speed of gravitational waves varies with time. Conversely, (Bassett *et al.* 2000) proposed a model in which $c_{photon}$ varies while $c_{gravity}$ is fixed. Therefore the Lorentz symmetry is "softly broken" (VSL theories often entail breaking Lorentz invariance, whereas dilaton theories do not), the "geometrical interpretation" is preserved, and the Bianchi identities are fulfilled. In particular, these "soft breaking" VSL scenarios are based on straightforward extensions of known physics, such as anomalous electromagnetic propagation in gravitational fields and so represent "minimalist" implementations of VSL theories. (Teyssandier 2002; 2004) showed that relativistic theories may be characterized in such a way that $c$ is not linked to any type of physical interaction and develops a theory of non-minimal coupling between electromagnetism and gravity in which the speed of light (as distinguished from $c$) varies. In our theory, the speed of light is a variable quantity and gravity has an electromagnetically induced nature though the speed of gravitational waves can be quite unrelated to the speed of light.

In view of this, we find that the possibility of a heterotic model (VSL plus inflation) is more attractive, even though our understanding of special relativity, general relativity and space-time will be significantly altered. However, the real significance of change in the fine structure is that such an anomaly necessarily indicates a change in one (VSL alone) or more of the quantities given in its definition (more probable), all of which are very important in determining the physical behavior of the universe. Assuming that a fundamental constant, say $c$, is in fact a function of time requires a close examination of all the other constants, here called concomitants that with $c$ enter in expressions that have a physical meaning. Even an invariant α would not exclude the possibility that these values might be changing in ratio to one another (Shea 2005). For instance, Bel (2003) in addition to the time dependency of *c*, derived the concomitant dependence of the electric permittivity $\epsilon$, the magnetic permeability μ, the unit of charge *e* and Plank's constant *h* under the assumption of the constancy of the fine structure constant α. Marciak-Kozlowska and Kozlowski have examined the possibility of both *c* and *ℏ* changing. Thus, in the equation for α discussed above, *ℏ* increases and *c* decreases so as to offset one another, allowing for an invariant fine structure constant even with variation in its parameters. One of the interesting consequences of Marciak-Kozlowska and Kozlowki's theory is that the universe becomes increasingly more quantum in nature as it ages.

As the last section should have made clear, the future of VSL is in the hands of observers. In such circumstances theorists are bound to continue with their musings.

4.2 Time varying dark components result in time varying α or the quantities given in its definition (invariant α), in either case VSL

Various models are proposed in literatures which join the variation of α to the evolution of the quintessence field (Anchordoqui & Goldberg 2003; Chiba & Kohri 2002; Wetterich 2003a, b; Dvali & Zaldarriaga 2002; Lee, Olive & Pospelov 2004; Olive & Pospelov 2002; Lee 2005; Bassett *et al.* 2000; Damour, Piazza & Veneziano 2002a, b; Barrow & Magueijo 2000). It is a characteristic feature of the quintessence scenario that fundamental coupling constants depend on time even in late cosmology where such a time variation could be observable and no such time dependence would be connected to dark energy if the latter occurs in the form of a cosmological constant (Chiba & Kohri 2002; Dvali & Zaldarriaga 2002; Wetterich 1988a,b, 2003b). Any field-theoretical description of a cosmological time variation of fundamental constants almost necessarily involves a scalar field which continues to evolve in time in the recent cosmological history. It is then quite natural (although not compulsory) to associate the potential and kinetic energy of this scalar field with quintessence (Wetterich 2003b).

If the recent observations suggesting a time variation of the fine structure constant within a 4D field theory are correct, they *necessarily* imply the existence of a time-dependent ultra-light scalar particle $\phi$ (Chiba & Kohri 2002; Dvali & Zaldarriaga 2002). This particle inevitably couples to nucleons through the α-dependence of their masses and thus mediates a composition-dependent "fifth force"-type long range interaction. Due to an interesting coincidence of the required time-scales (typically the onset of quintessence dominance suppresses variations in whatever 'varying' constant), the scalar filed in question can at the same time play the role of a quintessence field. If the field $\varphi$ driving the variation in α is a quintessence field, then its evolution is further constrained by observation. In particular, it must provide about 70% of the total energy density at present. The observed present value $\Omega_m \approx 0.3$ indicates that the effective "stop" in the quintessence field evolution should have happened rather recently - a value $z \approx 1$ for the drop in the variation rate of the fundamental couplings seems rather natural in this respect.

The evolution of the quintessence field, and therefore of α, is determined by the quintessence potential and such time variation can be explained for a wide class of quintessential potentials. Indeed, the quadratic coupling for the cosh-like potential (potentials similar to Sahni & Wang 2000) shows that the ratio $\Delta\alpha(z = 1.5)/ \Delta\alpha(z = 3)$ can be as small as 0.01 (see Figs. 4 and 5). This implies that in certain models the non-zero result of Webb *et al.* (2001) and Murphy, Webb & Flambaum (2003) for $\Delta\alpha/\alpha$ that spans a rather large range of redshifts and the better-sensitivity zero-result measurement of $\Delta\alpha/\alpha$ at one redshift $z = 1.5$ may not necessarily be contradictory (Lee, Olive & Pospelov 2004). In order to have a $O(10^{-5} - 10^{-6})$ relative change of α at redshifts $z \sim 1$, interestingly, these models typically require larger values of coupling between the scalar field and electromagnetic radiation.

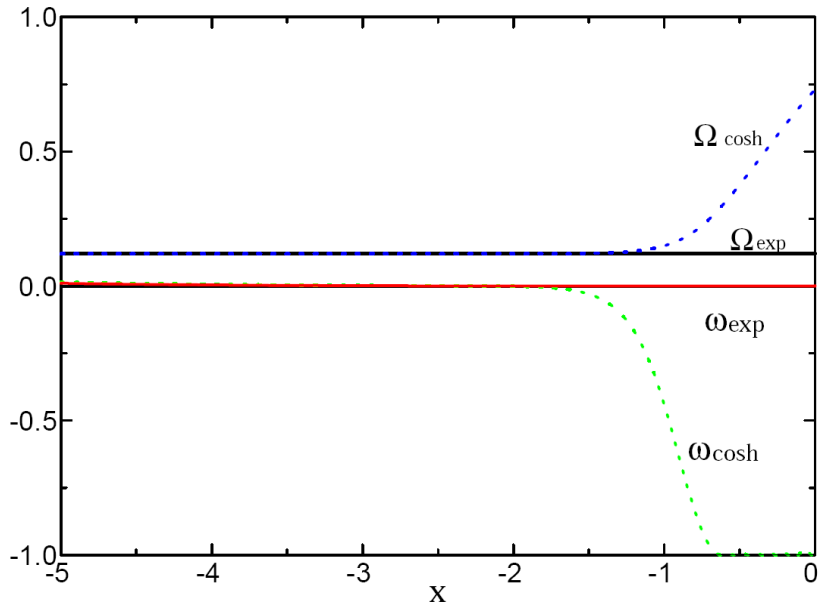

Figure 4. The evolution of $\Omega_\varphi$ and $w_\varphi$ for the simple exponential potential (solid) and the cosh potential (dotted). The main difference between two potentials appears at late times; (Lee, Olive & Pospelov 2004).

Anchordoqui & Goldberg (2003) have tied the variation of α to the evolution of the quintessence field proposed by (Albrecht & Skordis 2000):

$$V(\phi) = V_p(\phi)e^{-\lambda\phi} \text{ in which } V_p(\phi) = V_0(\phi - B)^\alpha + A; \qquad (8)$$

$$\rightarrow V(\phi) = V_0[(\phi - B)^\alpha + A]e^{-\lambda\phi} \qquad (9)$$

Interestingly, as mentioned above, Albrecht & Skordis (2000) is similar to the asymptotic form of Sahni & Wang (2000) in which $\tilde{V}_0$ is replaced by a $\phi$-dependent term $V_p(\phi)$. Both of them have analogous tracking properties with a minimum.

$$\frac{\Delta\alpha}{\alpha} \propto \frac{\Delta\phi}{M_{Pl}} \text{ (Anchordoqui \& Goldberg 2003; Chiba \& Kohri 2002)} \qquad (10)$$

The most general expansion of the function $\alpha(\phi)$ about its present day value $\alpha_0 = \alpha(\phi = \phi_{today})$ can be written as

$$\alpha = \alpha_0 + \lambda \frac{\phi}{M_{Pl}} + ... \text{ (Dvali \& Zaldarriaga 2002)} \qquad (11)$$

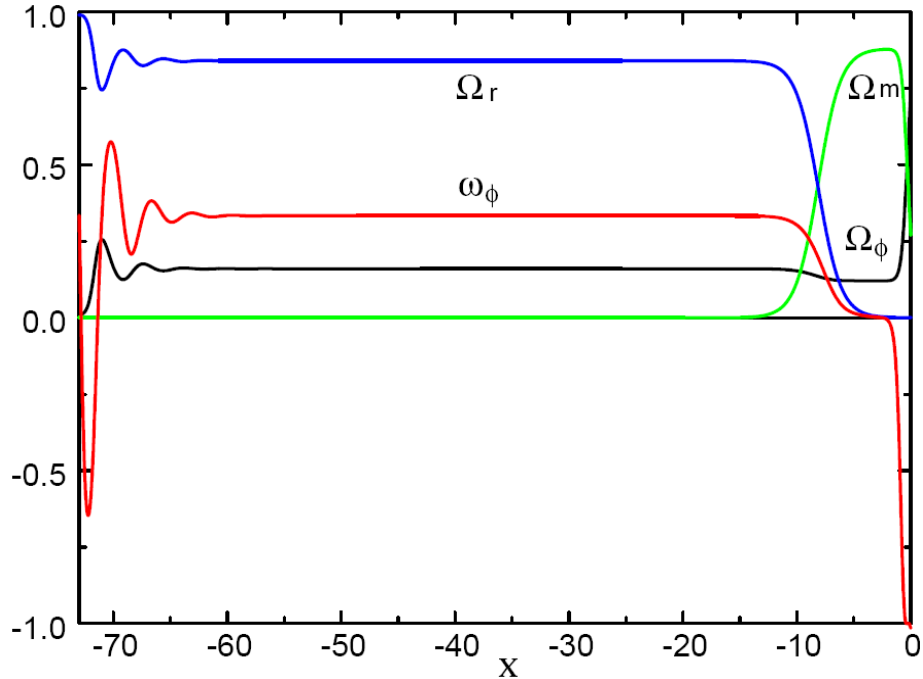

**Figure 5.** The cosh($\lambda\phi$) potential (Lee, Olive & Pospelov 2004). The cosmological evolution of the equation of state parameter, $w_\varphi$, and the energy density parameters, $\Omega_i$, of each component for $\lambda = 5$.

Crossover quintessence (CQ) (Wetterich 2003a) predicts that the time variation of fundamental couplings is substantially faster at redshift $z \approx 2$ than over the past few billion years. This could reconcile the reported time variation of the fine structure constant from quasar absorption lines with severe bounds from archeo-nuclear physics and high precision tests of the equivalence principle. The variation of α shows (through equation 10) that the motion of $\varphi$ as it comes to its present value is overdamped, so that the change in α is a monotonically decreasing function of $z$, therefore if there is some fluctuations, they can be because of the local spatial variation of matter density or Landau transitions. QSO measurements at larger redshifts should continue to show a monotonic decrease in Δα. This will allow for more precise restrictions on the shape of the function $α(z)$ which can be directly compared with the model predictions.

A small value of $|R| = \left|\dfrac{\Delta\alpha(z=0.13)}{\Delta\alpha(z=2)}\right|$ requires a quintessence model where the evolution of the scalar field has considerably slowed down in the recent history as compared to a redshift $z \approx 2$. In particular, a constant equation of state with $w$ independent of $z$ over a range $0 < z < 3$ will have severe difficulties to explain $R < 0.1$, also a bound $R < 1/50$ strongly favors quintessence with a time varying equation of state $w = p/\rho$, where the value of $(1+w)$ at present is substantially smaller than for $z = 2$ (Wetterich 2003a) and equivalently the velocity of sound in the medium is reduced (compressibility is increased) and this process has the main influence on the process of structure formation (transition to gravitational instability). On the level of the equation of state CQ corresponds to a crossover from a value $w \approx 1/3$ during the radiation dominated universe to a value $w < 1/3$ for $z < 0.5$. Wetterich (2003a) for a crossover quintessence predicts the fraction of quintessence at last scattering, to be 0.019 or 0.0083 for two CQ models.

In view of the bounds, coupling of quintessence is very small for ordinary matter - a larger coupling to dark matter remains possible, however (Wetterich 2003b). Chiba & Kohri (2002) explain the recent QSO data and the Oklo in terms of quintessence coupling to a photon which would require to a model with a large coupling between non-baryonic dark matter and the scalar field therefore consistent with our model, quintessence may couple to both non-baryonic dark matter and radiation (Olive & Pospelov 2002; Wetterich 2003a). Taking the reported time variation of $\alpha$ at face value would fix the coupling strength of the scalar field to matter and radiation (Wetterich 2003b). Because quintessence as a ultra-light scalar field would be dynamical and the exchange of light fields gives rise to long range forces, the interaction of the field with ordinary matter would result in the time variation of the constants of nature (like $\alpha$ or the ratio between the nucleon mass and the Planck mass) over cosmological time scales and the violation of the weak equivalence principle. The differential acceleration of two test bodies with equal mass can now be quantitatively related to the time variation of $\alpha$ in terms of observable cosmological parameters. If these QSOs observations prove valid, the proposed satellite experiment for testing the equivalence principle (STEP) should be able to detect the violation of the weak equivalence principle induced by a scalar force mediated by quintessence.

In a grand unified theory the time variation of $\alpha$ is directly related to a time variation of the strong gauge coupling and therefore also to a time variation of the nucleon mass $m_n$.

$$\frac{\partial \alpha}{\partial \ln \phi} = -6.7 \times 10^{-3} \frac{\partial \ln Z_F}{\partial \ln \phi} + 1.4 \times 10^{-5} \frac{\partial L_{Wn}}{\partial \ln \phi} \qquad (12)$$

More explicitly, we need to derive the rate of change of the cosmon field, $d\ln\phi/dt$ for a given model of quintessence. Then we can translate equation (12) into an equation for the time dependence of the fine structure constant and similar for other quantities (Wetterich 2003b).

$$\frac{\partial \alpha(z)}{\partial z} = \frac{\partial \alpha}{\partial \ln \phi} \frac{\partial \ln \phi}{\partial z} = \frac{\partial \alpha}{\partial \ln \phi} G(z) \qquad (13)$$

We will relate $G(z)$ to the equation of state of quintessence.

$$G(z) = \frac{\partial \ln \phi}{\partial z} = -\frac{1}{A + 4B} \left\{ \frac{3(1+w)}{1+z} - \frac{dw/dz}{1-w} \right\} \qquad (14)$$

The dependence of the fundamental constants on $z$ is not linear. Even for $w$ constant the rate of change of the couplings decreases for large $z$ proportional $(1 + z)^{-1}$. The change in $\alpha_{em}$ is reduced during cosmological epochs when $w$ decreases with decreasing $z$ (Wetterich 2003b).

Wetterich (2003b) observe that $\Omega_{quint}$ does not appear in the relation between $\phi$ and $z$ and therefore does not affect the time evolution of the fundamental couplings. The latter depends only on the equation of state $w$. Therefore according to Wetterich (2003b), time varying couplings are possible even if the scalar field plays no role in late cosmology, i.e. $\Omega_{quint} \ll 1$.

Already now, the reported QSO observation of $\Delta\alpha$, if confirmed, places important restrictions on quintessence. Models with a rather slow time evolution of the scalar field in a recent epoch ($z \leq 1$) and a more rapid evolution in the earlier universe ($z \geq 1$) are clearly favored (Wetterich 2003b).

4.3 Acceleration of universe expansion is imaginary and the negative pressure of quintessence is not obligatory

Until recently, there was strong prejudice against the idea that the Universe could be accelerating. There simply is no compelling theoretical framework that could accommodate an accelerating universe. Since the case for an accelerating universe continues to build, attempts have been made to improve the theoretical situation, with some modest success. Still, major open questions remain. All attempts to account for acceleration introduce a new type of matter (the "dark energy" or "quintessence") with an equation of state $p = w\rho$ ($w < 0$) relating pressure and energy density. Although we now know the present density of universe fairly accurately, we know precious little else about the dark energy.

The CMB has no sensitivity to evolution of $w$ (Frieman *et al.* 2003). CMB anisotropy alone cannot tightly constrain the properties of dark energy because of a strong degeneracy between the average equation-of-state and the matter density. Recent observations of high-redshift ($z > 0.3$) Type Ia supernovae (SNe Ia) provide the backbone of the body of evidence that we live in an accelerating universe (Riess et al. 2000a) whose content is dominated by vacuum energy (Riess et al. 1998; Perlmutter et al. 1999). The observations of supernovae in distant galaxies offer evidence for the detection of dark energy, under the assumption that distant and nearby supernovae are drawn from the same statistical sample (that is, that they are statistically similar enough for the purpose of this test). This is not at a very high level of statistical significance. There is no direct way to check this, and it is easy to imagine differences between distant and nearby supernovae of the same nominal type. More distant supernovae are seen in younger galaxies, because of the light travel time, and these younger galaxies tend to have more massive rapidly evolving stars with lower heavy element abundances. How do we know the properties of the supernovae are not also different? If SNe Ia are evolving, how are previous measurements of cosmological parameters from SNe Ia affected?

Also, the incomplete understanding of a number of astrophysical effects and processes (evolution, intergalactic dust, etc.) means that the supernovae results are preliminary and not yet definitive. Therefore, we have to consider that there may be a significant systematic error from differences between distant, high redshift, and nearby, low redshift, supernovae. I recommend Leibundgut's (2001) cautionary discussion of astrophysical uncertainties: the unknown nature of the trigger for the nuclear burning, the possibility that the Phillips correction to a fiducial luminosity actually depends on redshift or environment within a galaxy, and possible obscuration by intergalactic dust. There are also issues of physics that may affect this test (and others): the strengths of the gravitational or electromagnetic interactions may vary with time, and photon-axion conversion may reduce the number of photons reaching us. All of this is under active study (Peebles & Ratra 2003).

4.3.1 Time varying constants

As proposed in previous section, the main reason for the observed acceleration of universe expansion and the negative pressure of quintessence is the time varying constants, specially the variable speed of light (VSL). Recent VSL theories are firmly entrenched in the successes (and remaining failures) of the hot big bang theory of the universe. Observers in the VSL metric frame see the dimming of supernovae, because of the increase of the luminosity distance versus red shift, due to an increasing speed of light in the past universe, ~ 10% between $z = 0$ and $z \sim 2-3$. Moreover, in this frame, the scalar field $\phi$ describes a quintessence component in the Friedmann equation for the cosmic scale without acceleration.

Albrecht & Magueijo (1999) showed that $E \propto c^2$ for photons in free flight. They also showed that quantum mechanics remains unaffected by a changing $c$ if $\hbar \propto c$ (in the sense that quantum numbers are adiabatic invariants; a varying-$c$ implied a varying-$\hbar$. This connection has clear physical implications. But there is more: it could be that some of the strange properties of quantum mechanics—such as its non-locality- are the result of a form of 'faster than light' communication, Magueijo 2003). Then all relativistic energies scale like $c^2$. If for non-relativistic systems $\hbar \propto 1/c$, the Rydberg energy $E_R = m_e e^4 / (2\hbar^2)$ also scales like $c^2$. Hence all absorption lines, ignoring the fine structure, scale like $c^2$. When we compare lines from near and far systems we should therefore see no effects due to a varying $c$; the redshift $z$ is still $1 + z_e = a_o/a_e$, where o and e label epochs of observation and emission (Barrow & Magueijo 2000).

In order to examine luminosity distances, we need to reassess the concept of standard candles. For simplicity let us first treat them as black bodies. Then their temperature scales as $T \propto c^2$ (Albrecht & Magueijo 1999), their energy density scales as $\rho \propto T^4/(\hbar c)^3 \propto c^2$, and their emission power as $P = \rho/c \propto c$, implying that standard candles are brighter in the early universe if $\dot{c} < 0$. However, the power emitted by these candles, in free flight, scales like $c$; each photon's energy scales like $c^2$, its speed like $c$, and therefore its energy flux like $c$. The received flux, as a function of $c$, therefore scales like:

$$P_r = \frac{P_e c^2}{4\pi r^2 c} \propto c \qquad (15)$$

where $r$ is the conformal distance to the emitting object, and the subscripts $r$ and $e$ label received and emitted. In an expanding universe we therefore still have

$$P_r = \frac{P_{e0}}{4\pi r^2 a_0^2}\left(\frac{a}{a_0}\right)^2, \qquad (16)$$

where $P_{e0}$ is the emitting power of standard candles today. Notice that the above argument is still valid if the candles are not black bodies; it depends only on the scaling properties of emitted and received power.

We can now set up the Hubble diagram. Consider the Taylor expansion
$$a(t) = a_0[1 + H_0(t - t_0) - 1/2\, q_0 H_0^2\,(t - t_0)^2 + ...] \qquad (17)$$
where $H_0 = \frac{\dot{a}_0}{a_0}$ is the Hubble constant, and $q_0 = -\ddot{a}_0 a_0 / \dot{a}_0^2$ is the deceleration parameter. Hence up to second order $z = H_0\,(t_0 - t) + (1 + q_0/2)\,H_0^2\,(t - t_0)^2$, or
$$t_0 - t = 1/H_0\,[z - (1 + q_0/2)\,z^2 + ...]. \qquad (18)$$
From equation (16) we find that the luminosity distance $d_L$ is

$$d_L = \left(\frac{P_{e0}}{4\pi P_0}\right)^{1/2} = a_0^2\,\frac{r}{a} = a_0 r(1 + z_e). \qquad (19)$$

The conformal distance to the emitting object is given by $r = \int_t^{t_0} c(t)\,dt/a(t)$. From equation (17) we have that

$$r = c_0\,[(t_0 - t) + \frac{1-n}{2}\,H_0(t_0 - t)^2 + ...] \qquad (20)$$

where we have assumed that locally $c = c_0 a^n$ (that is $c = c_0[1 + nH_0(t - t_0) + ...]$).

Substituting (18) we finally have:

$$d_L = \frac{c_0}{H_0}\left[z + \frac{1}{2}(1 - (q_0 + n))z^2 + ...\right] \qquad (21)$$

(Had we assumed that $\hbar \propto c$ for all systems we would have got instead $d_L = (c_0/\tilde{H}_0)[z + 1/2(1 - (q_0(1 + 4n) + n))z^2]$, with $\tilde{H}_0 = (1 - 4n)H_0$. This does not affect any of the conclusions.)

We see that besides the direct effects of VSL upon the expansion rate of the universe, it also induces an effective acceleration in the Hubble diagram as an "optical illusion" (we are assuming that $c$ decreases in time: $n < 0$). This is easy to understand. We have seen that VSL introduces no intrinsic effects in the redshifting spectral line or in the dimming of standard candles with distance and expansion. The only effect VSL induces on the construction of the Hubble diagram is that for the same redshift (that is, the same distance into the past) objects are farther away from us because light travelled faster in the past (Barrow & Magueijo 2000).

4.3.2 Photon - pseudoscalar field conversion

Because the SNe observations probe length scales $l \sim H_0^{-1} \sim$ few $\times 10^3$ Mpc which are inaccessible to any particle physics experiments, it is natural to consider alternative explanations to the supernova data without cosmological dark energy. A simple such alternative is that light emitted by a distant supernova encounters an obstacle en route to us and gets partially absorbed. Recently it has been suggested that the dimness of high redshift supernovae is not due to the accelerated expansion of the universe, but rather due to mixing between the photons coming from these objects and a pseudo-scalar axion field in the intervening intergalactic plasma (Csaki, Kaloper & Terning 2002). This mechanism works only if the initial axion flux is much smaller than the initial photon flux. Grossman, Roy & Zupan (2002) find it likely that the initial axion flux is very small and therefore does not pose a problem to the CKT mechanism. This results in a luminosity-distance vs. redshift curve almost indistinguishable from that produced by the accelerating Universe, if the axion mass and coupling scale are $m \sim 10^{-16}$ eV, $M \sim 4 \cdot 10^{11}$ GeV. In our model, light is the boson dark matter (equivalent of axion dark matter) itself, therefore the above mechanism for dimming of supernovae is a probable one.

4.3.3 Systematic errors

High redshift (0.3 < $z$ <1.0) Type Ia supernovae are unexpectedly dim, a phenomenon readily attributed to a cosmological constant and an accelerating universe (Riess et al. 1998; Perlmutter et al. 1999). These cosmological conclusions rely on the *assumption* that SNe Ia have not evolved. Although systematic effects such as luminosity evolution, dimming by intervening extragalactic material (alternatively brightening due to gravitational lensing) continue to be a cause of some concern – recall that a luminosity evolution of ~ 25% over a lookback time of ~ 5 Gyr is sufficient to nullify the cosmological conclusions (Riess et al. 1999). (Riess et al. 1999) has stressed that we cannot directly determine the impact of the apparent evolution on determinations of cosmological parameters (Sahni 2004).

Sources of systematic error from K-corrections, extinction, selection effects, Malmquist bias and evolution are investigated in literature, and their effects estimated.

Selection effects can introduce systematic errors as a function of redshift, as can uncertain K-corrections, or an evolution of the SN Ia progenitor population as a function of look-back time. These effects, if they are large, will limit our ability to measure the ($D_L$, $z$) relation accurately, and have the potential to sap the potency of high-redshift SN Ia for measuring cosmological parameters.

Additional complications arise because magnitude-limited selection might prevent extinguished objects from being discovered with equal efficiency in both nearby and distant samples, or because the smaller angular size of distant objects makes it more difficult to detect SN deep inside spirals — those likely to be extinguished. In short, it would not be surprising to have systematic differences between the average extinctions of objects discovered in the nearby and distant supernova searches.

The systematic uncertainty in translating between filter systems limits the accuracy with which we are able to measure luminosity distances at high redshift.

It is a very difficult task to ensure systematic errors in (B − V) < 0.02 mag in either the nearby or distant sample, and we believe that this uncertainty will be our largest source of systematic error.

It has been pointed out by Kantowski, Vaughan, & Branch (1995) that large-scale structure could magnify (or demagnify) a SN's light through weak gravitational lensing as it travels to an observer. Given the size of other systematic errors, uncertainties due to gravitational lensing are not likely to be of major concern up to $z \approx 1$.

The data presented by Garnavich et al. (1998) and Riess et al. (1998) indicate that we can measure the distances to high-z supernovae with a statistical uncertainty of σ = 0.2 mag (10%) per object. With only 10 objects a comparison of z ≈ 0 and z ≈ 0.5 can be made to a precision of better than 5% – leaving systematic uncertainties as a major contributor to the total error budget. A summary of the contributions to high-redshift supernova distance uncertainties is given in Table 1. This table shows that programs to measure cosmology will most likely be limited by the possibility of the evolution of SN Ia explosions with look-back time. Future work to address this possible problem will be as important as obtaining large numbers of objects at high redshift.

Table 1. Summary of Error Contributions to High Supernova Distances (Schmidt et al. 1998)

| Systematic Uncertainties(1σ) | (mag) | Statistical Uncertainties | (mag) |
|---|---|---|---|
| Photometric System Zero Point[a] | 0.05 | Individual Zero Points | 0.02 |
| Selection Effects | 0.02 | Shot noise | 0.15 |
| Evolution | < 0.17 | K-corrections | 0.03 |
| Evolution of Extinction Law | 0.02 | Extinction | 0.1 |
| Gravitational Lensing | 0.02 | σ of SN Ia | 0.15 |

[a]Includes propogated effect on extinction, 3.1 σ $E(B-V)$.

With a single object it is difficult to make serious conclusions about cosmological parameters, regardless of the distance precision it offers, because there is no way to judge systematic errors in an empirical way.

4.3.4 Intergalactic dust and Evolution

The two sources most likely to obscure distant SNe Ia and affect interpretation of them are dust (Aguirre 1999a, 1999b; Totani & Kobayashi 1999) and evolution. Although, Riess et al. (2000a) observations disfavor a 30% opacity of SN Ia visual light by dust as an alternative to an accelerating universe.

Rest-frame evolution is the potential pitfall in using high-$z$ SNe Ia to measure the cosmological parameters. The lack of a complete theoretical model of SNe Ia including the identification of their progenitor systems makes it difficult to access the expected evolution between $z = 0$ and 0.5 (Livio 2000; Umeda et al. 1999; Höflich, Wheeler, & Thielemann 1998). An impressive degree of similarity has been observed between the spectral and photometric properties of nearby and high-z SNe Ia (Schmidt et al. 1998; Perlmutter et al. 1998, 1999; Riess et al. 1998; Filippenko et al.; but see also Riess et al. 1999; Drell, Loredo, & Wassermann 2000).

However, it is not known what kind or degree of change in the observable properties of SNe Ia would indicate a change in the expected peak luminosity by 30%. The detection of such a change would cast doubt on the reliability of the luminosity distances from high-z SNe Ia (Riess et al. 2000a). A continued failure to measure any difference between SNe Ia near and far would increase the confidence (although never prove) that evolution does not contaminate the cosmological measurements from high-z SNe Ia. We expect dimming from systematic effects to grow larger with increasing redshift (Coil et al. 2000). Obtaining data for SNe at $z \geq 0.8$ is therefore critical for testing systematic bias versus real cosmological effects and the presence of a nonzero $\Omega_\Lambda$.

4.3.5 Fluid dynamics of the cosmic foam (large coherent velocity flows)

Type Ia supernovae have recently been employed to measure the peculiar motions of galaxies. The great distances to which SNe Ia can be seen make them particularly well-suited to constraining large-scale velocity flows (Riess 2000b).

(Riess 2000b) reviewed the use of SNe Ia to measure the motion of the Local Group (include 44 SNe Ia). Hudson believes that most of the bulk motion of the 405 km s$^{-1}$ which is required to agree with the predicted motion of the Local Group is due to sources beyond 8000 km s$^{-1}$. Using 28 primarily photographically observed SNe Ia, Miller & Branch (1992) were able to discern the gravitational influence of the Virgo cluster, i.e. Virgocentric infall. Because the average depth of their sample was only about $cz$=2000 km s$^{-1}$, their analysis was insensitive to the motion of the Local Group and the influence of the Great Attractor.

Riess in an analysis in his thesis (Riess, Press, & Kirshner 1995) of 13 new SNe Ia from the Calan/Tololo Search made the first detection of the motion of the Local Group using SNe Ia. The sample had an effective depth of $cz$ =7000 km s$^{-1}$ and a typical distance precision of 6%.

The predicted peculiar velocities of SNe Ia are a function of the local mass in the Universe $\Omega_M$ as well as the degree to which the positions of galaxies indicate the location of mass (i.e., the bias parameter).

Riess (2000b) mentioned an analysis by Zehavi, Riess, Kirshner & Dekel (1998) which gives a marginal indication of a so-called "Hubble Bubble". From 44 SNe Ia, Zehavi *et al.* (1998) found an indication at the 2-3$\sigma$ confidence level of a local excess expansion of 6% within 7000 km s$^{-1}$. This increase in the global Hubble expansion appears to be compensated by a small decrease beyond this depth after which the Hubble expansion appears to settle to its global value. The model proposed by the authors is that we may live within a local void bounded by a wall or density contrast at ~100 Mpc. More SNe Ia (and other distance indicators) will be required to test this provocative result.

We live in a universe which one of the largest clustered structures like Great Attractor are largely affected by the fluid dynamics of the cosmic foam (Mathewson & Ford 1994), therefore, it is probable that the observed acceleration of the SNe Ia expansion be *real* (with respect to us) but because of the dynamical evolution of the cosmic foam itself, without the need to exotic and strikingly unusual theoretical framework that could accommodate an accelerating universe.

In VSL theories, the link between luminosity distance and look-back time is obviously modified: with a higher $c$ in the past, objects with the same look-back time are further away (Moffat 2003b; Barrow & Magueijo 2000; Belinchon 2000; Alvarez & Bel 1973). If, however, one takes into account the QSO observations (Webb *et al.* 1999), one finds that any corrections to the construction of the Hubble diagram, because of VSL alone, must be small (Barrow & Magueijo 2000; Magueijo 2003), unless as mentioned above, in addition to the VSL, variations in some other quantities (like $e$, $G$ or $\hbar$) (Bel 2003; Magueijo 2003) and/or the influence of dynamical evolution of the cosmic foam are also introduced. This might enable one to consider models where the extra energy component of the universe is not a cosmological constant or quintessence with negative pressure but a fluid with a different equation of state.

4.4 The 'speed of light' *c* is not a fundamental constant

The 'speed of light' $c$ - the maximum attainable speed of the low-energy quasiparticles like phonons or rotons (quantized vortices) - is not a fundamental constant, but a material parameter. Since the 'speed of light' is not a fundamental quantity, but is determined by the material parameters of the liquid, it cannot enter explicitly into any physical result or equation written in covariant form. The *'speed of light' c(n)* depends on $n = N/V$, the particle density, in the vacuum. However, the 'speed of light' $c$ becomes the fundamental constant of the effective theory arising in the low-energy corner. The structure of the quasiparticle spectrum in superfluid $^4$He becomes more and more universal the lower the energy. Inspired by superfluid $^4$He, in the low-energy limit of superfluid boson dark matter, one obtains the linear spectrum, $E(p, n) \rightarrow c(n)p$, which characterizes the phonon modes - quanta of sound waves. Their spectrum depends only on the 'fundamental constant', the speed of 'light' $c(n)$ obeying

$$c^2(n) = (n/m)(d^2\varepsilon/dn^2), \tag{22}$$

$m$ of the particle as a microscopic parameter, $\varepsilon(n)$ vacuum energy density;

All other information on the microscopic particle nature of the liquid is lost. Since phonons have long wave-length and low frequency, their dynamics is within the responsibility of the effective theory.

At sufficient low temperature, the phonon excitations are the more numerous and dominate the thermal properties, whereas at higher temperatures the number of rotons increases exponentially with temperature, and rotons become predominant thermodynamically therefore we have mainly roton radiation against phonon radiation. The effective theory is unable to describe the high-energy part of the spectrum - the rotons - which can be determined in a fully microscopic theory only (Volovik 2003). Onsager (1949) and Feynman (1955) found that $\nabla \times v_s \neq 0$ at singular lines, the quantized vortices. Penrose's twistor is a vortex ring, as is a magnetic field. It is interesting to note that vortex rings can sustain transverse vibrations (analogous to guitar string vibration). The subject of vortices and the equations which apply to them were fully calculated by Kelvin (1880) and lead to exactly the correct equations for electromagnetism as explained by Stowe (1996). Kelvin proved mathematically that linear disturbances in a saturated 3D vortex fluid (he termed a vortex sponge) would produce propagation of pure transverse waves identical to the equations and properties that describe the propagation of light through space. It also is interesting to note that Maxwell used this conceptual model as the basis for his derivation of the EM relationships (Mingst and Stowe). Mingst and Stowe infer the superfluidity of the aether from presence of the vortex rings. Winterberg (1992, 2002) proposed that vacuum aether should be modeled as phonon-roton superfluid and the roton hypothesis can explain both of the CDM and quintessence. The Planck aether is superfluid, and it can without expenditure of energy form a tangle of quantized vortex filaments permitting the transmission of two types of waves: One associated with a symmetric displacement of the vortex lattice leading to gravitational waves, and one associated with an antisymmetric displacement leading to electromagnetic waves.

The ground states of the quantum bosonic and fermionic liquids correspond to the vacuum. The low-energy bosonic and fermionic quasiparticles in the quantum liquids correspond to matter. The low-energy modes with linear spectrum $E = cp$ can be described by the relativistic type effective theory. The speed $c$ of sound plays the role of the speed of light. This 'speed of light' is the 'fundamental constant' for the effective theory. As mentioned in previous section, if 'speed of light' $c$ be a decreasing function of time with a decelerated rate and the 'speed of light' is in fact the speed $c$ of sound in superfluid boson dark matter therefore the 'speed of light' depends only on the equation of state $w$ and the quasiparticle spectrum of superfluid dark matter should bents upwards en route for high energy.

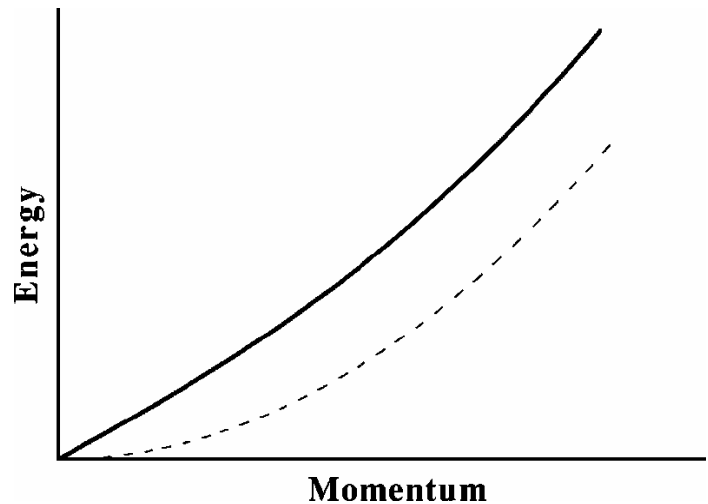

Figure 6. The dispersion relation of an elementary excitation in the weakly interacting Bose gas, here the photon fluid (Chiao & Boyce 1999).

In the mid-nineteenth century, the mathematical concept of the field was developed to describe the apparent continuity of matter, light, and gravity. A field has a value at each point in space, in contrast to the properties of a particle which are localized to a tiny region of space. Note that fields appeared in reductionist physics over a century ago. Pressure and density are two examples of matter fields. In continuous elastic media, pressure and density propagate as sound waves when the media are excited. As the phenomena of electricity and magnetism became better understood, they were also described in terms of fields. When Maxwell discovered that the equations which united electricity with magnetism called for the propagation of electromagnetic waves in a vacuum at the speed of light, it was suggested that the vacuum was not empty but filled with an elastic medium - the aether - whose excitation produced the phenomenon of light (Stenger 1992). The resemblance between elastodynamics and electromagnetism has been known for a long time, and was first pointed out by Sir William Thomson. The mechanical version of aether in classical physics was introduced by Christian Huygens in order to explain the transmission of light via compression waves, analogous to the transmission of sound in air. In this form Aether became the potential mediator of all mysterious 'action at a distance forces' such as gravity, magnetism and electricity. Karlsen (1998) proposed that by redefining some of the terms used in elastodynamics in an elastic medium of infinite extension, one can readily show that Navier-Stokes equation can be brought over to a form that exactly matches that of Maxwell's electromagnetic equations even when the equations are written out in a four-dimensional manifold.

Superfluid dark matter is converted to normal fluid at a heat source, flows away from it and, in giving up its entropy to the heat sink, is converted back to superfluid again; As $T \to 0$, a small heat pulse produces a group of virtually non-interacting phonons (or photons) which travel ballistically at velocity of first sound between the source and sink.

When an external observer measures the propagation of 'light' (sound, or other massless low-energy quasiparticles), he or she finds that the speed of light is coordinate-dependent. Moreover, it is anisotropic: for instance, it depends on the direction of propagation with respect to the flow of the superfluid vacuum. On the contrary, the inner observer always finds that the 'speed of light' (the maximum attainable speed for low-energy quasiparticles) is an invariant quantity. This observer does not know that this invariance is the result of the flexibility of the clocks and rods made of quasiparticles: the physical Lorentz-Fitzgerald contraction of length of such a rod and the physical Lorentz slowing down of such a clock (the time dilation) conspire to produce an effective special relativity emerging in the low-energy corner. These physical effects experienced by low-energy instruments do not allow the inner observer to measure the 'aether drift', i.e. the motion of the superfluid vacuum: the Michelson-Morley-type measurements of the speed of massless quasiparticles in moving 'aether' would give a negative result. The low-energy rods and clocks also follow the anisotropy of the vacuum and thus cannot record this anisotropy. As a result, all the inner observers would agree that the speed of light is the fundamental constant. Living in the low-energy corner, they are unable to believe that in the broader world the external observer finds that, say, in superfluid boson dark matter 'speed of light' varies, depending on the direction of propagation. The invariance of the speed of sound in inhomogeneous, anisotropic and moving liquid as measured by a local inner observer is very similar to the invariance of the speed of light in special and general relativity. In the same manner, the invariance of the speed holds only if the measurement is purely local. If the measurement is extended to distances at which the gradients of $c$ and $v_s$ become important, the measured speed of light differs from its local value.

Therefore factorization of $\alpha$ in terms of the dimensionful parameters - electric charge $e$ and speed of light $c$ - would be artificial, especially if one must choose between two considerably different speeds of 'light', $c_\perp$ and $c_\parallel$. Such a factorization can be appropriate for an inner observer, for whom the 'speeds of light' in different directions are indistinguishable, but not for an external observer, for whom they are different.

The simplest interpretation and most correct, of Michelson-Morley is that - light is the aether or superfluid dark matter. Both signal and carrier waves are the aether (Twain 1997). Nature is continuous and self-connected and the aether is not a gas but a quantum liquid. The aether is light and plasma - with an insignificant amount of gas, liquid and solid thrown in as contaminant. The quantum liquid vigorously churned by a strong shear flow, mainly from expansion itself, in the dark ages and a cosmic foam of 'contaminants' as large scale structure forms on the surface of the 'ocean of light'. This picture is the brief history of the universe.

When, some seventy years after Einstein's prediction of the existence of a Bose-Einstein condensate (BEC), the first BEC gas was prepared from Rb vapour in 1995, project co-director and Nobel Prize winner Eric Cornell remarked: "This state could never have existed naturally anywhere in the universe. So the sample in our lab is the only chunk of this stuff in the universe..." Investigations of the nature of dark matter have led us to the diametrically opposite conclusion, viz. that a BEC may well be the most abundant form of matter in the cosmos– and a viable solution to the problem of "missing mass'' and even quintessence. Therefore, this statement is the most inexperienced sentences in the whole science history!

ACKNOWLEDGMENT
I thank Sepehr Arbabi Bidgoli for wide-ranging discussions.